\documentclass[lettersize,journal]{IEEEtran}
\usepackage{amssymb,amsmath,latexsym,amsfonts,amsthm,mathtools}
\usepackage{algorithmic}
\usepackage{algorithm}
\usepackage{array}
\usepackage{textcomp}
\usepackage{stfloats}
\usepackage{float,subcaption}
\usepackage{url}
\usepackage{verbatim}
\usepackage{graphicx}
\usepackage[hyperref]{xcolor}
\usepackage{hyperref}
\hypersetup{colorlinks, breaklinks, citecolor=blue, linkcolor=blue, urlcolor=blue}
\usepackage[nameinlink,noabbrev,sort&compress]{cleveref}
\usepackage{cite}
\Crefname{equation}{Eq.}{Eqs.}
\Crefname{figure}{Fig.}{Figs.}
\theoremstyle{plain}

\newtheorem*{problem}{Problem}
\theoremstyle{remark}
\newtheorem{remark}{Remark}
\newtheorem{assumption}{Assumption}

\newtheorem{definition}{Definition}
\theoremstyle{definition}

\newcommand{\sign}{\text{sign}}
\newcommand{\diag}{\text{diag}}
\IEEEoverridecommandlockouts

\begin{document}

\title{Cooperative Guidance and Control for Active Asset Protection with Time-Varying Agent Speeds}

\author{Ram Milan Kumar Verma, Shashi Ranjan Kumar,~\IEEEmembership{IEEE Senior Member}, and  Hemendra Arya
  \thanks{R. M. K. Verma, S. R. Kumar, and H. Arya are with the Department of Aerospace Engineering, Indian Institute of Technology Bombay, Powai, Mumbai 400076, India. (\tt{Emails}: rmverma@aero.iitb.ac.in, srk@aero.iitb.ac.in, arya@aero.iitb.ac.in)}}
  
\markboth{Journal of \LaTeX\ Class Files,~Vol.~XX, No.~XX, Month~Year}%
{Shell \MakeLowercase{\textit{et al.}}: A Sample Article Using IEEEtran.cls for IEEE Journals}

\maketitle

\begin{abstract}
    Protecting an asset against threats is a challenging problem in an era of continuously evolving intelligent attacks. This requires cooperation between the asset and the defender to share information and jointly maneuver. 
    To address this problem, this work proposes a cooperative guidance and control strategy for active asset protection against a maneuvering threat. This work develops a joint maneuver strategy where both the defender and the asset coordinate their time-varying speeds and courses to neutralize/capture the attacker. 
    The control strategy is formulated around three coupled geometric and temporal objectives. The first objective is to set the line-of-sight rate between the asset and the attacker to zero, putting the attacker on a collision course and reducing their maneuvering. The second objective is to maintain the defender on the line-of-sight between the asset and the attacker. This ensures that the attacker faces the defender first before reaching the vicinity of the asset. Lastly, the defender is also guided to pursue the attacker based on the time-to-go estimates between the defender and the attacker. While keeping these objectives in mind, the control actions for the asset and the defender are jointly designed, fostering cooperation between the two. 
    The stability of the proposed strategy is established using a Lyapunov-based approach. Numerical simulations performed show the effectiveness of the proposed cooperative strategy in ensuring the successful capture of a maneuvering threat.
\end{abstract}


\section{Introduction}
\IEEEPARstart{I}{n} recent years, rapid technological advancements in autonomous vehicles have led to their use across multiple domains, ranging from maritime surveillance and orbital servicing to territorial defense. Protection of a high-value or indispensable asset is critical and central to many of these domains. In ever-changing, emerging scenarios, intelligent attackers are using cheaper, agile, autonomous agents. This requires the asset to also deploy a more agile defender.  Central to these operations is the old pursuit-evasion problem, a dynamic conflict between one or more pursuers attempting to capture an evader while the evader strives to avoid capture. The complexity and challenges are evolving, requiring new algorithms to address them. Modern scenarios demand that simply launching a defensive agent is not enough, and a joint maneuver might be required to capture the incoming threat.

The asset could be airborne, on the water surface, underwater, or on the ground. For instance, in active aircraft protection, it may launch a defender interceptor to intercept an incoming attacker, creating a cooperative team. Similarly, in maritime environments, a high-value ship (which could be carrying expensive cargo or humans) can be protected against an attacking ship using small autonomous, uncrewed surface vessels (USVs) that are faster and highly maneuverable \cite{10.1109/TCNS.2025.3649094,10.1016/j.oceaneng.2022.112742,10.1109/TCYB.2019.2958548}. This problem involving three agents: the asset, the defender, and the attacker, is also being applied to space domains \cite{10.1109/TAES.2020.2998197}, such as for the orbital servicing \cite{10.1007/s40295-025-00501-x}.

In the existing literature, the problem has been approached using diverse methodologies, including differential game, optimal control, geometric formation control, nonlinear guidance laws, and learning-based approaches \cite{10.1016/j.automatica.2025.112629,10.1016/j.oceaneng.2026.124314,10.1007/s40295-025-00501-x,10.2514/6.2010-7876,10.2514/6.2021-1881,10.1109/JOE.2025.3634663}. Initial efforts in this direction were made in \cite{10.1109/TAES.1976.308338} in the case of defense against torpedo or interceptor attack against a ship, in which they derived kinematic relations for three bodies under the approximation of constant-bearing trajectories. In \cite{10.2514/1.G000659}, authors presented a triangle intercept guidance law, in which if the defender maintains being on the line-of-sight (LOS) joining attacker and target (asset), then the attacker will have to intercept the defender first before getting close to the asset. Another LOS-based cooperative strategy was presented in \cite{10.1109/TAES.2020.3046328}, which used a cost function based on $\ell_p$-norm for static optimization at each instant to derive the control inputs. 

Differential game theory \cite{10.1016/j.oceaneng.2026.124314,10.2514/1.G004068} is another approach to derive such guidance strategies for three-body engagements. The solutions to the game between two adversarial teams were obtained by numerically solving the Riccati equation. In \cite{10.1109/TAES.2011.5751240}, the authors studied the game of defending an asset under linear quadratic formulation. In \cite{10.2514/1.61832}, the authors derived the capture conditions for the pursuer to capture the evader. Optimization theories \cite{10.3182/20110828-6-IT-1002.02587} were also employed to develop guidance strategies for either defense or attack while successfully evading the defender. The cooperative strategy presented in \cite{10.2514/1.G001083} aimed to maximize the separation between the attacker and the target by formulating the solution as a two-point boundary-value problem. However, this is a computationally more intensive approach. Also, the guidance strategies in \cite{10.2514/1.61832,10.2514/1.G001083, 10.1109/TAES.2011.5751240} were obtained under a linearized setting, which may not be effective for scenarios with larger deviations from the operating conditions. 

As the cooperative strategies evolved, the natural question arises about the levels of cooperation. There can be varying levels of cooperation between the asset and the defender. The cooperation can be at the information-sharing level or in joint maneuvering. The authors in \cite{10.2514/6.2012-4908} presented guidance strategies for one-way and two-way cooperation between the asset and the defender. 
In \cite{10.1007/s10846-022-01570-y}, a time-constrained interception strategy was presented for an evader-defender team. The authors in \cite{10.1109/LCSYS.2020.3041799} presented a multi-defender strategy based on a cooperative salvo with a leader defender, while the evader lures the attacker by nullifying its LOS rate.  The authors in \cite{10.2514/6.2021-1881} utilized a proportional navigation strategy by the defender and the pursuer in 3D Cartesian space. 
Ample work has been done in the interceptor guidance literature, but most of it assumes that the attacker, defender, and target are moving at constant speeds. The interceptors cannot change speed because they lack an active axial propulsion system. However, for USVs, this is not the case, and a surge thruster can be used to vary speed. The authors in \cite{10.1109/TCNS.2025.3649094} present a cooperative capture strategy of a faster evader using multiple defender USVs. This used a technique based on an Apollonius circle, where an encircling task and a hunting task were designed to form an encirclement formation and approach the faster-evading USV. A similar technique was employed in \cite{10.1016/j.oceaneng.2022.112742}, with swarms of defenders and collision-avoidance capabilities. The authors in \cite{10.1109/TCYB.2019.2958548} presented a distributed pursuit algorithm to encircle and capture a freely moving faster evader. However, it was not always possible to deploy multiple defender USVs to protect against the incoming threat. The authors in \cite{10.1007/s10846-016-0379-3} presented an escape strategy for an evader with holonomic constraints against multiple pursuers. Another interesting work inspired by biological predator-prey dynamics, \cite{10.1016/j.automatica.2025.112629} proposed an ``Alert-Turn'' strategy for unicycle robots. Their work also modeled the trade-off between speed and maneuverability observed in nature. The focus and objectives vary depending on the domain of applications. For some applications, interception time is critical, and for some applications, the operation needs to be performed with minimal energy consumption. In the case of space applications, as in \cite{10.1007/s40295-025-00501-x}, an orbital pursuit-evasion strategy was presented that focused on the consumed energy rather than the interception time. 

Building on these foundations in literature, the following are the key contributions of the current work.
\begin{itemize}
    \item In this work, we propose a cooperative guidance and control strategy for joint maneuvering between the asset and the defender against an incoming threat. Compared to existing works \cite{10.2514/6.2010-7876,10.1109/TAES.2025.3649483,10.1109/TCYB.2019.2958548,10.1109/LCSYS.2020.3041799} in the interceptor guidance literature, which assume agents move at constant speeds, the proposed work accounts for the time-varying speeds of the asset, the defender, and the attacker. Unlike interceptors, USVs can actively change speed over time and steer their course. 
    \item To foster strategic collaboration between the defender and the asset, the design goals are set in terms of coupled geometric and temporal objectives.
    First, maneuver so that the defender maintains itself on the LOS between the asset and the attacker. This forces the attacker to face the defender first before it can reach the asset. 
    Second, ensure that the asset maneuvers to nullify the LOS rate between the attacker and the asset. 
    By doing this, the attacker might be deceived into entering a collision course and might decide to maneuver less. Taking advantage of this, the defender, guided by a true proportional navigation (TPN)-based strategy, intercepts the attacker at a specified, anticipated time.
    \item The developed strategy accounts for nonlinear engagement kinematics to describe the relative motion between the three agents, thereby ensuring good performance across a broader operational envelope as opposed to \cite{10.2514/1.61832,10.2514/1.G001083, 10.1109/TAES.2011.5751240}. 
    \item The proposed strategy employs a control law that incorporates both radial and tangential accelerations of the asset and the defender. Apart from enhancing the agility and cooperation, this formulation provides the necessary flexibility for precise trajectory modulation and reliable interception of the maneuvering threat. That is, if the asset control inputs are insufficient (which might be the case, as the asset has lower maneuverability than the defender), the defender, taking advantage of its higher maneuverability, performs the necessary actions to capture the attacker. 
    \item The framework also gives flexibility to operate on varying levels of control authority (for example, in cases of failure of speed of control or steering control) to achieve coordination between the asset and the defender team. 
\end{itemize}

The idea is to protect the asset by teaming up with the defender against a maneuvering attacker by setting geometric and temporal goals. A significant deviation from the prior work is the consideration of time-varying speeds of all the agents involved. The temporal objective is set according to time-to-go estimates to intercept the attacker using TPN-based guidance law. Note that the idea of time-varying speeds can be directly utilized by the USVs to protect an asset ship, as the USVs and the ship can control their speeds as well as steer their heading. 

The rest of the paper is organized as follows: In \Cref{sec:Problem}, the kinematics of the relative motion between the three agents is described, following which a formal description of the problem addressed is presented. The design strategy and main results are derived in \Cref{sec:MainResults}. The proposed strategy is validated via MATLAB\textsuperscript{\textregistered} simulation in \Cref{sec:simulation} and conclusions are drawn in \Cref{sec:conclusion}.
\section{Problem Formulation} \label{sec:Problem}

In this section, we first define the reference frames necessary to describe the motion of the agents involved. With respect to the defined reference frames, the engagement kinematics of the relative motion among the asset, the defender, and the attacker are presented.
Following this, the proposed strategy will be derived. The motion of the agents is described with respect to the Earth-fixed frame denoted by $OX_IY_I$, whose $x$-axis is towards North and $y$-axis is towards East. Note that the subscript ($S$) will be used for the asset ship, the subscript ($D$) for the defender USV, and the subscript ($A$) for the attacker. 

\subsection{Engagement kinematics}
\begin{figure}[h!]
    \centering
    \includegraphics[width=1\linewidth]{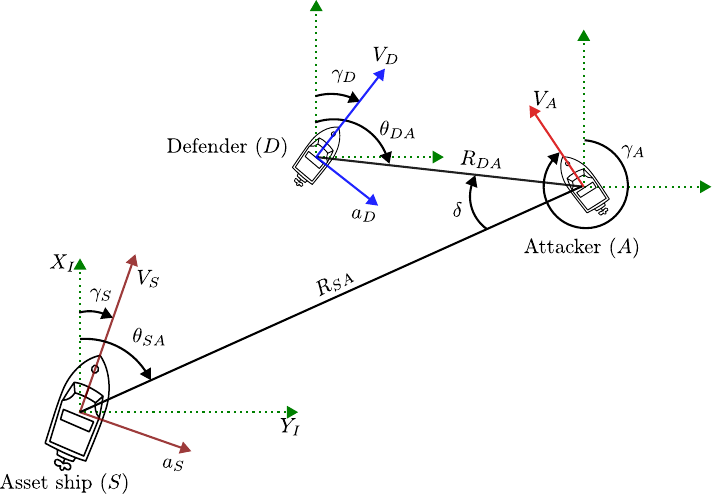}
    \caption{Representation of the relative motion scenario between the asset, the defender, and the attacker.}
    \label{fig:3body_engagement}
\end{figure}

In this subsection, we describe the equations of motion among the agents in polar coordinates. Consider the scenario depicted in \Cref{fig:3body_engagement}. Let the time-varying speeds of the asset ship, the defender USV, and the attacker be denoted by $V_S(t)$, $V_D(t)$, and $V_A(t)$, respectively.
For brevity, we drop the argument $t$, and it is understood that the speeds are time-varying, unless stated otherwise. The range and LOS angle between the asset ship and the attacker are denoted by $R_{SA}$ and $\theta_{SA}$, respectively. Similarly, the range and the LOS angle between the defender USV and the attacker are denoted by $R_{DA}$ and $\theta_{DA}$, respectively. The course angles (measured clockwise from the North) of the asset ship, the defender, and the attacker are denoted by $\gamma_S$, $\gamma_D$, and $\gamma_A$, respectively. In terms of range and LOS variables, the relative motion between defender and attacker can be described by 
\begin{align}
    \dot{R}_{DA} =V_A \cos(\gamma_A - \theta_{DA}) - V_D \cos(\gamma_D-\theta_{DA}) =V_{R_{DA}}\label{eqn:Rdot_DA},\\
    R_{DA}\dot{\theta}_{DA} = V_A \sin(\gamma_A - \theta_{DA}) - V_D \sin(\gamma_D-\theta_{DA})=V_{\theta_{DA}}.\label{eqn:Rtheta_dot_DA}
\end{align}
\Cref{eqn:Rdot_DA} describes the evolution of the closing distance between the defender and the attacker, while \Cref{eqn:Rtheta_dot_DA} describes the component of the relative velocity perpendicular to the LOS between the defender and the attacker. Similarly, the relative motion between the asset ship and the attacker is governed by 
\begin{subequations}
\begin{align}
     \dot{R}_{SA} &=V_A \cos(\gamma_A - \theta_{SA}) - V_S \cos(\gamma_S-\theta_{SA}) \label{eqn:Rdot_SA},\\
    R_{SA}\dot{\theta}_{SA} &= V_A \sin(\gamma_A - \theta_{SA}) - V_S \sin(\gamma_S-\theta_{SA}).\label{eqn:Rtheta_dot_SA}
\end{align}
\end{subequations}
The variations of the course angles for the ship, defender, and attacker are given by
\begin{align}
    \dot{\gamma}_S &= \dfrac{a_S}{V_S}, \quad \dot{\gamma}_D = \dfrac{a_D }{V_D}, \quad \dot{\gamma}_A = \dfrac{a_A}{V_A}, \label{eqn:gamma_dot}
\end{align}
where lateral accelerations of the agents are given by $a_{i}~\forall~i\in \{S,D,A\}$. Note that the radial accelerations are directly related to the change in the speeds of the agents and can be denoted by $\dot{V}_i~\forall~i\in \{S, D, A\}$. Before we formally define the problem addressed in this work, we revisit the required definitions and remarks and state a few assumptions.

\begin{definition}[Time-to-go]\label{defn:tgo}
    Time-to-go for a defender-attacker engagement is defined as the time taken by the defender to capture the attacker. For a defender USV guided by TPN-based guidance law, the time-to-go can be expressed as (derived in \cite{ghose1994,kumar2022true})
    \begin{equation}
        t_{\text{go}} = -\dfrac{R(V_{R}+2\lambda)}{V_{R}^2 + V_{\theta}^2 + 2\lambda V_{R}}, \label{eqn:tgo-nonmanuevering}
    \end{equation}
    where $\lambda$ is a guidance parameter for TPN, $R$ denotes the range, and $V_R$ and $V_\theta$ are the components of relative velocity along and perpendicular to the LOS.
\end{definition}

Note that the time-to-go expression in \Cref{eqn:tgo-nonmanuevering} was obtained for a defender against a non-maneuvering attacker, as it is not tractable to obtain a closed-form expression against a maneuvering attacker. However, the same expression may serve as a time-to-go estimate against a maneuvering attacker, assuming a constant heading angle at a given time instant. We now define the error in time-to-go as
\begin{equation}\label{eqn:et_}
    e(t) = t_{\text{go}} - t_{\text{go}}^d, \quad t_{\text{go}}^d = T_d - t_{\text{el}},
\end{equation}
where $T_d$ is the feasible terminal time in which the defender should neutralize the attacker, $t_{\text{go}}^d$ is the desired time-to-go at a time instant $t_{\text{el}}$ (elapsed time). 

\begin{remark}
    Note from \Cref{eqn:et_} that as time progresses, if we maintain time-to-go error to zero, then $t_{\rm go}\rightarrow t_{\text{go}}^d$. Also, $t_{\text{go}}^d\to 0$ as $t_{\text{el}}\to T_d$. From \Cref{eqn:tgo-nonmanuevering}, it can be concluded that as $R\to 0$, $t_{\rm go}\to 0$. Another condition when $t_{go}$ becomes zero is when $V_R = -2 \lambda$. Therefore, by suitably choosing $\lambda$, this situation can be avoided. Hence, nullifying the error in time-to-go is a necessary and sufficient condition for nullifying the range, which implies interception at the desired terminal time.
\end{remark}
\begin{assumption}
    The asset ship has lower maneuvering capability, while the defender USV and the attacker are matched in their capabilities. 
\end{assumption}
\begin{assumption}
    All three agents involved are assumed to be point mass vehicles, and we only consider kinematics into consideration for guidance design purposes.
\end{assumption}
As the proposed strategy uses geometric and temporal objectives, we need to define another quantity, $\delta$, as the difference between the LOS angles of the defender-attacker and the asset-attacker engagement, given by
\begin{align}
    \delta = \theta_{DA} - \theta_{SA}.\label{eqn:delta}
\end{align}
From \Cref{fig:3body_engagement}, it follows that $\delta$ denotes the angle between the two lines-of-sight $SA$ (the line joining the asset and the attacker) and the $DA$ (the line joining the defender and the attacker). Under the stated assumptions (both assumptions represent viable scenarios and do not pose a stringent challenge to their real applicability), we now present the problem statement addressed in this work. 

\begin{problem}
Consider the engagement scenario between the asset, the defender, and the attacker as depicted in \Cref{fig:3body_engagement}. Given the engagement kinematics relation between the defender and attacker as in \Cref{eqn:Rdot_DA,eqn:Rtheta_dot_DA}, and between the asset ship and the attacker as in \Cref{eqn:Rdot_SA,eqn:Rtheta_dot_SA}, the objective of guidance design is to cooperatively steer the asset ship and the defender such that the attacker gets captured by the defender before it reaches the vicinity of the asset ship. Mathematically, the objective is to simultaneously design the control inputs for the asset ship ($\dot{V}_S$ and $a_S$) and the defender ($\dot{V}_D$ and $a_D$) such that the following three objectives are met :
\begin{enumerate}
    \item $\dot{\theta}_{SA} \rightarrow 0$: This will engage the attacker to be on a collision course with the asset, thereby making it nonmaneuvering and making it easier for the defender to capture the attacker.
    \item $\delta \rightarrow 0$: The defender USV will be on the LOS joining the asset ship and the attacker. This will ensure that the attacker has to encounter the defender first before it reaches the vicinity of the asset.
    \item $e(t)\rightarrow 0$: Defender will steer its trajectory to not only be on asset-attacker LOS but also approach the attacker for a successful interception.
\end{enumerate}
\end{problem}

\begin{remark}
    A notable feature of the proposed strategy is that if the specified time constraint, $T_d$, is not met, then placing the defender on the LOS between the asset and the attacker ensures that the attacker will have to face the defender first.
\end{remark}
We design guidance strategies without performing any linearizations, thereby proposed strategies remain applicable for engagements with even larger initial deviations. 

\section{Main results}\label{sec:MainResults}
In this section, we first obtain the necessary error dynamics in terms of the available control inputs for the asset ship and the defender USV. Subsequently, we present rigorous mathematical proofs of stability for the proposed strategy. Our objective is to devise a strategy to drive the error variables $e_\delta$, $e(t)$, and $\dot{\theta}_{SA}$ to zero. We now proceed to obtain the dynamics of the LOS between the ship and the attacker, as well as the defender and the attacker.

\subsection{Dynamics of LOS rate $\dot{\theta}_{SA}$}
On differentiating \Cref{eqn:Rtheta_dot_SA} with respect to time, one can obtain
\begin{align}
    R_{SA}&\ddot{\theta}_{SA} + \dot{R}_{SA}\dot{\theta}_{SA} = \dot{V}_A \sin(\gamma_A - \theta_{SA}) \nonumber \\
    & - \left[ \dot{V}_S \sin(\gamma_S - \theta_{SA}) + V_S \cos(\gamma_S - \theta_{SA})(\dot{\gamma}_S - \dot{\theta}_{SA}) \right] \nonumber \\
    &+ V_A \cos(\gamma_A - \theta_{SA})(\dot{\gamma}_A - \dot{\theta}_{SA}) .\label{eqn:Rtheta_ddot_SA1}
\end{align}
On rearranging terms in \Cref{eqn:Rtheta_ddot_SA1} and using \Cref{eqn:Rdot_DA,eqn:gamma_dot}, the LOS dynamics between the ship and the attacker is obtained as 
\begin{equation}
\begin{split}
    \ddot{\theta}_{SA}  &= - \dfrac{2\dot{R}_{SA}\dot{\theta}_{SA}}{R_{SA}} \\
    &\quad - \dfrac{ \dot{V}_S \sin(\gamma_S - \theta_{SA}) + a_S \cos(\gamma_S - \theta_{SA})}{R_{SA}} \\
    &\quad + \dfrac{ \dot{V}_A \sin(\gamma_A - \theta_{SA}) + a_A \cos(\gamma_A - \theta_{SA})}{R_{SA}}. \label{eqn:theta_ddot_SA}
\end{split}
\end{equation}
We can now rewrite \Cref{eqn:theta_ddot_SA} in a compact form as 
\begin{align}
    \ddot{\theta}_{SA}  &= f_{\theta_{SA}} + g_{\theta_{SA1}}\dot{V}_S + g_{\theta_{SA2}}a_S,\label{eqn:theta_ddot_SA_compact}
\end{align}
where the terms $f_{\theta_{SA}}$, $g_{\theta_{SA1}}$, and $g_{\theta_{SA2}}$ are defined as 
\begin{align*}
    f_{\theta_{SA}} &= - \dfrac{2\dot{R}_{SA}\dot{\theta}_{SA}}{R_{SA}} \\
    &\quad + \dfrac{ \dot{V}_A \sin(\gamma_A - \theta_{SA}) + a_A \cos(\gamma_A - \theta_{SA})}{R_{SA}},  \\
    g_{\theta_{SA1}} &= - \dfrac{ \sin(\gamma_S - \theta_{SA}) }{R_{SA}}, \quad 
    g_{\theta_{SA2}} = - \dfrac{ \cos(\gamma_S - \theta_{SA})}{R_{SA}}.
\end{align*}

\subsection{Dynamics of angle $\delta$}
Next, we obtain the dynamics of angle $\delta$. On differentiating \Cref{eqn:delta} with respect to time, we get
\begin{align}
    \dot{\delta} = \dot{\theta}_{DA} - \dot{\theta}_{SA}.
\end{align}
From \Cref{eqn:Rtheta_dot_DA,eqn:Rtheta_dot_SA}, we see that the control input terms do not appear in the expression of $\dot{\delta}$. We further differentiate \Cref{eqn:delta} with respect to time to obtain
\begin{align}
    \ddot{\delta} = \ddot{\theta}_{DA} - \ddot{\theta}_{SA} \label{eqn:delta_ddot}
\end{align}
On taking time differentiation of \Cref{eqn:Rtheta_dot_DA}

\begin{align}
    \dot{R}_{DA}\dot{\theta}_{DA} &+ R_{DA}\ddot{\theta}_{DA} = V_A (\dot{\gamma}_A - \dot{\theta}_{DA}) \cos(\gamma_A - \theta_{DA})  \nonumber \\
    &+ \dot{V}_A \sin(\gamma_A - \theta_{DA}) - \dot{V}_D \sin(\gamma_D - \theta_{DA}) \nonumber \\
    &  - V_D (\dot{\gamma}_D - \dot{\theta}_{DA}) \cos(\gamma_D - \theta_{DA}). \label{eqn:Rtheta_ddot_DA}
\end{align}
Grouping all the terms involving $\dot{\theta}_{DA}$ together, we can rewrite 
\begin{align}
    &R_{DA}\ddot{\theta}_{DA} + \dot{R}_{DA}\dot{\theta}_{DA} \nonumber\\
    &=  \dot{V}_A \sin(\gamma_A - \theta_{DA}) + V_A \dot{\gamma}_A \cos(\gamma_A - \theta_{DA})  \nonumber \\
    & -  \dot{V}_D \sin(\gamma_D - \theta_{DA}) + V_D \dot{\gamma}_D \cos(\gamma_D - \theta_{DA})  \nonumber \\
    & -\dot{\theta}_{DA}\left[V_A \cos(\gamma_A - \theta_{DA}) - V_D \cos(\gamma_D-\theta_{DA})\right].
    \label{eqn:Rtheta_ddot_DA1}
\end{align}
Now, using \Cref{eqn:Rdot_DA,eqn:gamma_dot} into \Cref{eqn:Rtheta_ddot_DA1}, one may obtain
\begin{align}
    R_{DA}\ddot{\theta}_{DA} &+ 2\dot{R}_{DA}\dot{\theta}_{DA} =  \dot{V}_A \sin(\gamma_A - \theta_{DA}) \nonumber \\
    &+ a_A \cos(\gamma_A - \theta_{DA}) -  \dot{V}_D \sin(\gamma_D - \theta_{DA}) \nonumber \\
    &-  a_D \cos(\gamma_D - \theta_{DA}). \label{eqn:Rtheta_ddot_DA2}
\end{align}
Rearranging \Cref{eqn:Rtheta_ddot_DA2} leads to
\begin{equation}
\begin{split}
    \ddot{\theta}_{DA}  &= - \dfrac{2\dot{R}_{DA}\dot{\theta}_{DA}}{R_{DA}} \\
    &\quad - \dfrac{ \dot{V}_D \sin(\gamma_D - \theta_{DA}) +  a_D \cos(\gamma_D - \theta_{DA}) }{R_{DA}} \\
    &\quad + \dfrac{ \dot{V}_A \sin(\gamma_A - \theta_{DA}) + a_A \cos(\gamma_A - \theta_{DA})}{R_{DA}}.
    \label{eqn:theta_ddot_DA}
\end{split}
\end{equation}

Using \Cref{eqn:theta_ddot_DA,eqn:theta_ddot_SA} into \Cref{eqn:delta_ddot}, we get
\begin{equation}
\begin{split}
    \ddot{\delta} &= - \dfrac{2\dot{R}_{DA}\dot{\theta}_{DA}}{R_{DA}} + \dfrac{2\dot{R}_{SA}\dot{\theta}_{SA}}{R_{SA}} \\
    &\quad - \dfrac{\dot{V}_D \sin(\gamma_D - \theta_{DA}) +  a_D \cos(\gamma_D - \theta_{DA})}{R_{DA}} \\
    &\quad + \dfrac{ \dot{V}_S \sin(\gamma_S - \theta_{SA}) + a_S \cos(\gamma_S - \theta_{SA})}{R_{SA}} \\
    &\quad + \dfrac{ \dot{V}_A \sin(\gamma_A - \theta_{DA}) + a_A \cos(\gamma_A - \theta_{DA})}{R_{DA}} \\
    &\quad - \dfrac{ \dot{V}_A \sin(\gamma_A - \theta_{SA}) + a_A \cos(\gamma_A - \theta_{SA})}{R_{SA}}. \label{eqn:deltaddot}
\end{split}
\end{equation}
Note that the \Cref{eqn:deltaddot} contains the control terms, namely, $\dot{V}_S$,  $a_S$, $\dot{V}_D$, and $a_D$. Therefore, the relative degree of angle $\delta$ with respect to the control inputs is two. For the ease of representation, \Cref{eqn:deltaddot} can be rewritten as
\begin{align}
    \ddot{\delta} = f_\delta + g_{\delta_1}\dot{V}_S + g_{\delta_2}a_S + g_{\delta_3}a_D + g_{\delta_4}\dot{V}_D, \label{eqn:deltaddot_compact}
\end{align}
where the terms $f_\delta,~ g_{\delta_1},~ g_{\delta_2},~ g_{\delta_3}$, and $g_{\delta_4}$ are defined as
\begin{align*}
    f_\delta &= - \dfrac{2\dot{R}_{DA}\dot{\theta}_{DA}}{R_{DA}} + \dfrac{2\dot{R}_{SA}\dot{\theta}_{SA}}{R_{SA}} \\
    &\quad + \dfrac{ \dot{V}_A \sin(\gamma_A - \theta_{DA}) + a_A \cos(\gamma_A - \theta_{DA})}{R_{DA}} \\
    &\quad - \dfrac{ \dot{V}_A \sin(\gamma_A - \theta_{SA}) + a_A \cos(\gamma_A - \theta_{SA})}{R_{SA}}, \\
    g_{\delta_1} &= \dfrac{\sin(\gamma_S - \theta_{SA})}{R_{SA}}, \quad
    g_{\delta_2} = \dfrac{\cos(\gamma_S - \theta_{SA})}{R_{SA}}, \\
    g_{\delta_3} &= -\dfrac{\cos(\gamma_D - \theta_{DA})}{R_{DA}}, \quad 
    g_{\delta_4} = -\dfrac{\sin(\gamma_D - \theta_{DA})}{R_{DA}}.
\end{align*}
Next, we derive the dynamics of error in time-to-go between the defender and the attacker.

\subsection{Dynamics of impact time error $e(t)$}
To ensure the defender pursues the attacker and captures it within a given time, we use a time-to-go-based strategy. For the defender and attacker engagement scenario as depicted in \Cref{fig:3body_engagement}, the time-to-go expression (refer \Cref{defn:tgo}) can be written as
\begin{equation}\label{eqn:tgo}
    t_{\text{go}} = -\dfrac{R_{DA}(V_{R_{DA}}+2\lambda)}{V_{R_{DA}}^2 + V_{\theta_{DA}}^2 + 2\lambda V_{R_{DA}}}.
\end{equation}
It is evident from \Cref{eqn:tgo} that by controlling the components of relative velocity, time-to-go can be altered to meet the terminal time constraint. However, it must still be ensured that the time-to-go converges to zero, guiding the defender USV to the attacker. The necessity and sufficiency condition for the same was established in \cite{kumar2022true}, which states that if the $\lambda$ is chosen to meet the condition $\lambda \gg V_D/2$, then the convergence of $t_{\text{go}}$ (given by \Cref{eqn:tgo}) to zero is necessary and sufficient condition for the defender USV to intercept the attacker. The error dynamics can be obtained by differentiating the \Cref{eqn:et_} with respect to time and is given by
\begin{align}\label{eqn:et_dot}
    \dot{e} = \dot{t}_{\rm{go}} - \dot{t}^d_{\rm{go}} = \dot{t}_{\rm{go}} + 1.
\end{align}
To obtain the derivative $\dot{t}_{\rm{go}}$, we differentiate \Cref{eqn:tgo} with respect to time, which yields 
\begin{equation}
\begin{split}
    \dot{t}_{\text{go}} 
    &= -\frac{\left(V_{R_{DA}}^2 + 2\lambda V_{R_{DA}} + R_{DA}\dot{V}_{R_{DA}}\right)}{V_{R_{DA}}^2 + V_{\theta_{DA}}^2 + 2\lambda V_{R_{DA}}} \\
    &\quad + \frac{ 2 R_{DA}(V_{R_{DA}}+2\lambda)\big[(V_{R_{DA}} + \lambda)\dot{V}_{R_{DA}} + V_{\theta_{DA}}\dot{V}_{\theta_{DA}}\big]}{\left(V_{R_{DA}}^2 + V_{\theta_{DA}}^2 + 2\lambda V_{R_{DA}}\right)^2}. \label{eqn:tgodot}
\end{split}
\end{equation}
Note that \Cref{eqn:tgodot} contains the time derivatives of $V_{R_{DA}}$ and $V_{\theta_{DA}}$.
By differentiating \Cref{eqn:Rdot_DA} with respect to time, we get 
\begin{equation}
\begin{split}
    \dot{V}_{R_{DA}} &= - V_A \sin(\gamma_A - \theta_{DA})(\dot{\gamma}_A - \dot{\theta}_{DA})  \\
    &\quad  \dot{V}_A \cos(\gamma_A - \theta_{DA})  - \dot{V}_D \cos(\gamma_D - \theta_{DA}) \\
    &\quad + V_D \sin(\gamma_D - \theta_{DA})(\dot{\gamma}_D - \dot{\theta}_{DA}),
\end{split}
\end{equation}
which upon using \Cref{eqn:gamma_dot} leads to
\begin{equation}
\begin{split}
    \dot{V}_{R_{DA}} 
    &= \dot{V}_A \cos(\gamma_A - \theta_{DA}) - a_A \sin(\gamma_A - \theta_{DA}) \\
    &\quad + \dot{\theta}_{DA} \left[ V_A \sin(\gamma_A - \theta_{DA}) - V_D \sin(\gamma_D - \theta_{DA}) \right] \\
    &\quad - \dot{V}_D \cos(\gamma_D - \theta_{DA}) + a_D \sin(\gamma_D - \theta_{DA}). \label{eqn:VRDA_dot}
\end{split}
\end{equation}
Substituting expression for $V_{\theta_{DA}}$ from \Cref{eqn:Rtheta_dot_DA} into \Cref{eqn:VRDA_dot}, we get
\begin{equation}
\begin{split}
    \dot{V}_{R_{DA}} &= \dot{V}_A \cos(\gamma_A - \theta_{DA}) - a_A \sin(\gamma_A - \theta_{DA}) \\
    &\quad + \dot{\theta}_{DA}V_{\theta_{DA}} - \dot{V}_D \cos(\gamma_D - \theta_{DA}) \\
    &\quad + a_D \sin(\gamma_D - \theta_{DA}).\label{eqn:VRDA_dot1}
\end{split}
\end{equation}
We further obtain the expression for $\dot{V}_{\theta_{DA}}$ by differentiating \Cref{eqn:Rtheta_dot_DA} with respect to time as
\begin{equation}
\begin{split}
    \dot{V}_{\theta_{DA}} &= V_A \cos(\gamma_A - \theta_{DA})(\dot{\gamma}_A - \dot{\theta}_{DA}) \\
    &\quad +  \dot{V}_A \sin(\gamma_A - \theta_{DA}) - \dot{V}_D \sin(\gamma_D - \theta_{DA}) \\
    &\quad - V_D \cos(\gamma_D - \theta_{DA})(\dot{\gamma}_D - \dot{\theta}_{DA}).
\end{split}
\end{equation}
Using \Cref{eqn:gamma_dot}, one can obtain
\begin{equation}
\begin{split}
    \dot{V}_{\theta_{DA}} &= \dot{V}_A \sin(\gamma_A - \theta_{DA}) + a_A \cos(\gamma_A - \theta_{DA}) \\
    &\quad - \dot{V}_D \sin(\gamma_D - \theta_{DA}) - a_D \cos(\gamma_D - \theta_{DA}) \\
    &\quad - \dot{\theta}_{DA} \left[ V_A \cos(\gamma_A - \theta_{DA}) - V_D \cos(\gamma_D - \theta_{DA}) \right].\label{eqn:VthetaDA_dot}
\end{split}
\end{equation}
Substituting $V_{R_{DA}}$ from \Cref{eqn:Rdot_DA} into \Cref{eqn:VthetaDA_dot}, results in
\begin{equation}
\begin{split}
    \dot{V}_{\theta_{DA}} &= \dot{V}_A \sin(\gamma_A - \theta_{DA}) + a_A \cos(\gamma_A - \theta_{DA}) \\
    &\quad - \dot{V}_D \sin(\gamma_D - \theta_{DA}) - a_D \cos(\gamma_D - \theta_{DA}) \\
    &\quad - \dot{\theta}_{DA}V_{R_{DA}}. \label{eqn:VthetaDA_dot1}
\end{split}
\end{equation}
Now, on substituting for $\dot{V}_{R_{DA}}$ and $\dot{V}_{\theta_{DA}}$ from \Cref{eqn:VRDA_dot1,eqn:VthetaDA_dot1}, respectively, into $\dot{t}_{\rm{go}}$ expression in \Cref{eqn:tgodot} and performing some algebraic simplifications, we get
\begin{align}
    \dot{t}_{\text{go}} &= -1 + \underbrace{{f}_t + g_{t_1} \dot{V}_A + g_{t_2} a_A}_{F_t} + g_{t_3} a_D + g_{t_4} \dot{V}_D, \label{eqn:tgodot_compact}
\end{align}
where the terms $ {f}_t, ~g_{t_1}, ~g_{t_2}, ~g_{t_3}$, and $g_{t_4}$ are defined as
\begin{align*}
    {f}_t &= \dfrac{2\lambda(V_{R_{DA}} + 2\lambda)V_{\theta_{DA}}^2}{(V_{R_{DA}}^2 + V_{\theta_{DA}}^2 + 2\lambda V_{R_{DA}})^2},\\
    g_{t_1} &= \frac{\left(V_{R_{DA}}^2 - V_{\theta_{DA}}^2 +4 \lambda V_{R_{DA}} + 4\lambda^2\right)R_{DA} \cos(\gamma_A - \theta_{DA})}{\left(V_{R_{DA}}^2 + V_{\theta_{DA}}^2 + 2\lambda V_{R_{DA}}\right)^2} \\
    &\quad + \frac{2 (V_{R_{DA}}+2\lambda) V_{\theta_{DA}}  R_{DA}\sin(\gamma_A - \theta_{DA})}{\left(V_{R_{DA}}^2 + V_{\theta_{DA}}^2 + 2\lambda V_{R_{DA}}\right)^2},\\
     g_{t_2} &=  \frac{\left(V_{\theta_{DA}}^2 - V_{R_{DA}}^2 -4 \lambda V_{R_{DA}} - 4\lambda^2\right)R_{DA} \sin(\gamma_A - \theta_{DA})}{\left(V_{R_{DA}}^2 + V_{\theta_{DA}}^2 + 2\lambda V_{R_{DA}}\right)^2} \\
    &\quad + \frac{2 (V_{R_{DA}}+2\lambda) V_{\theta_{DA}}  R_{DA}\cos(\gamma_A - \theta_{DA})}{\left(V_{R_{DA}}^2 + V_{\theta_{DA}}^2 + 2\lambda V_{R_{DA}}\right)^2},\\
    g_{t_3} &= \frac{R_{DA} \sin(\gamma_D - \theta_{DA}) \big[ (V_{R_{DA}} + 2\lambda)^2 - V_{\theta_{DA}}^2 \big]}{(V_{R_{DA}}^2 + V_{\theta_{DA}}^2 + 2\lambda V_{R_{DA}})^2} \\
    &\quad + \frac{R_{DA} \sin(\gamma_D - \theta_{DA}) \big[ 2V_{\theta_{DA}}(V_{R_{DA}} + 2\lambda) \big]}{(V_{R_{DA}}^2 + V_{\theta_{DA}}^2 + 2\lambda V_{R_{DA}})^2} ,\\ 
    g_{t_4} &= -\frac{R_{DA} \cos(\gamma_D - \theta_{DA}) \big[ (V_{R_{DA}} + 2\lambda)^2 - V_{\theta_{DA}}^2 \big]}{(V_{R_{DA}}^2 + V_{\theta_{DA}}^2 + 2\lambda V_{R_{DA}})^2} \\
    &\quad - \frac{R_{DA} \cos(\gamma_D - \theta_{DA}) \big[ 2V_{\theta_{DA}}(V_{R_{DA}} + 2\lambda) \big]}{(V_{R_{DA}}^2 + V_{\theta_{DA}}^2 + 2\lambda V_{R_{DA}})^2}.
\end{align*}
Using \Cref{eqn:et_dot,eqn:tgodot_compact}, we get the dynamics of time-to-go error as 
\begin{align}
    \dot{e} = F_t + g_{t_3} a_D + g_{t_4} \dot{V}_D,~F_t = {f}_t + g_{t_1} \dot{V}_A + g_{t_2} a_A.\label{eqn:edot_compact}
\end{align}
With all these error dynamics in place, we now derive the control commands to meet the set geometric and temporal objectives to achieve the mission goal of intercepting the attacker by the defender. 

\subsection{Derivation of control command}
To design the guidance and control command, consider the switching surfaces defined as 
\begin{align}
    \mathcal{S} = \begin{bmatrix}
        \mathcal{S}_\delta \\ \mathcal{S}_t \\ \mathcal{S}_{\theta_{SA}}
    \end{bmatrix} = \begin{bmatrix}
        \dot{\delta} + k_\delta \delta \\ e \\ \dot{\theta}_{SA}
    \end{bmatrix}. \label{eqn:S}
\end{align}
On differentiating \Cref{eqn:S} with respect to time yields
\begin{align}
    \dot{\mathcal{S}} = \begin{bmatrix}
        \ddot{\delta} + k_\delta \dot{\delta} & \dot{e} & \ddot{\theta}_{SA}
    \end{bmatrix}^\top.\label{eqn:Sdot}
\end{align}
Using \Cref{eqn:deltaddot_compact,eqn:edot_compact,eqn:theta_ddot_SA_compact} into \Cref{eqn:Sdot}, we get
\begin{align}
    \dot{\mathcal{S}} &= \begin{bmatrix}
        f_\delta + g_{\delta_1}\dot{V}_S + g_{\delta_2}a_S + g_{\delta_3}a_D + g_{\delta_4}\dot{V}_D  + k_\delta \dot{\delta} \\
        F_t + g_{t_3}a_D + g_{t_4}\dot{V}_D\\
         f_{\theta_{SA}} + g_{\theta_{SA1}}\dot{V}_S + g_{\theta_{SA2}}a_S
    \end{bmatrix},
\end{align}
which can be rewritten in the matrix-vector form as
\begin{align}
    \dot{\mathcal{S}} &= \mathcal{F} + \mathcal{G}\mathcal{U}, \quad \mathcal{U}= \begin{bmatrix}
        \dot{V}_S & a_S & a_D & \dot{V}_D
    \end{bmatrix}^\top, \label{eqn:Sdot_compact}
\end{align}
where the terms $\mathcal{F}$ and $\mathcal{G}$ are defined as 
\begin{align}
    \mathcal{F} = \begin{bmatrix}
f_\delta + k_\delta \dot{\delta} \\
F_t \\
f_{\theta_{SA}}
\end{bmatrix}, ~~ 
\mathcal{G} = \begin{bmatrix}
g_{\delta_1} & g_{\delta_2} & g_{\delta_3} & g_{\delta_4} \\
0 & 0 & g_{t_3} & g_{t_4}\\
g_{\theta_{SA1}} & g_{\theta_{SA2}} & 0 & 0
\end{bmatrix}.\label{eqn:FG}
\end{align}
Note that the control input vector, $\mathcal{U}$, is composed of linear accelerations ($\dot{V}_S$ and $\dot{V}_D$) and lateral accelerations ($a_S$ and $a_D$) of the asset and the defender.
\begin{remark}\label{rem:ctrl_strategy}
    The proposed guidance framework is adaptable to various asset maneuverability constraints. Depending on the asset ship's maneuverability, we can either change only the asset ship's speed, steer the asset ship's heading while keeping the speed constant, or vary both. 
\end{remark}
Consider a Lyapunov function candidate in terms of sliding surfaces as given by  
\begin{align}
    \mathcal{W} = \dfrac{1}{2}\mathcal{S}^\top \mathcal{S}. \label{eqn:W}
\end{align}
On differentiating \Cref{eqn:W} with respect to time and using \Cref{eqn:Sdot_compact}, we get
\begin{align}
    \dot{\mathcal{W}} = \mathcal{S}^\top \dot{\mathcal{S}} =\mathcal{S}^\top \left[\mathcal{F} + \mathcal{G}\mathcal{U}\right]. \label{eqn:Wdot}
\end{align}
Choosing the control input $\mathcal{U}$ as
\begin{align}
    \mathcal{U} = \mathcal{G}^{\dagger}\left[ -\mathcal{F} - \mathcal{M}\sign(\mathcal{S})\right], \label{eqn:U}
\end{align}
where $\mathcal{M} = \diag [\mathcal{M}_1~~\mathcal{M}_2~~\mathcal{M}_3]$ with $\mathcal{M}_i >0$ for $i\in \{1,2,3\}$, and $\mathcal{G}^{\dagger}$ represents the pseudo inverse ($ \mathcal{G}^{\dagger} = \mathcal{G}^\top (\mathcal{G} \mathcal{G}^\top)^{-1} $). Using \Cref{eqn:U,eqn:Wdot}, we obtain $\dot{\mathcal{W}} = -\mathcal{M}_1|\mathcal{S}_1| -\mathcal{M}_2|\mathcal{S}_2| - \mathcal{M}_3|\mathcal{S}_3| <0$. The negative definiteness of $\dot{\mathcal{W}}$ ensures that the sliding surface converges to zero, leading to the convergence of errors as well.

\begin{figure}[h!]
    \centering
    \includegraphics[width=\linewidth]{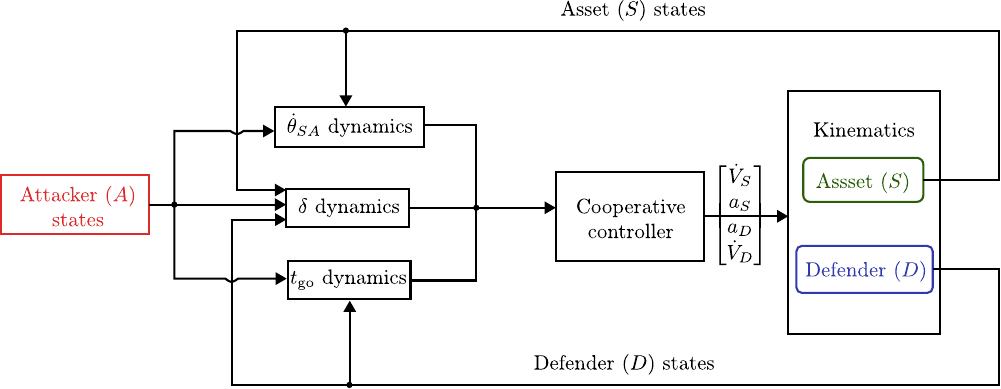}
    \caption{Working of cooperative control strategy.}
    \label{fig:control_schematic}
\end{figure}

The schematic of the cooperative control strategy is depicted in \Cref{fig:control_schematic}.
Next, we consider different scenarios (as outlined in \Cref{rem:ctrl_strategy}) in which we analyze the operation of the control methodology with varying levels of control authority available to the asset ship and the defender.

\subsection{Asset ship maneuvering ($a_S \neq 0$) with constant speed ($\dot{V}_S=0$)}
This represents an engagement scenario in which the asset ship can steer to change its direction but does not control its speed. The asset ship can continue moving with some constant speed. It can also choose to evade with maximum speed to give the defender more time to capture the attacker. The speed control of the asset ship is not utilized, that is,  $\dot{V}_S=0$. This reduces the control matrix $\mathcal{G}$ in \Cref{eqn:FG} as 
\begin{align}
    \mathcal{G} = \begin{bmatrix}
 g_{\delta_2} & g_{\delta_3} & g_{\delta_4} \\
 0 & g_{t_3} & g_{t_4}\\
 g_{\theta_{SA2}} & 0 & 0
\end{bmatrix}.\label{eqn:G_aS}
\end{align}
Using the expressions of $g_{\theta_{SA2}},~g_{\delta_3},~ g_{t_4}, ~g_{\delta_4}$, and $g_{t_3}$ as derived earlier in \Cref{eqn:theta_ddot_SA_compact,eqn:deltaddot_compact,eqn:tgodot_compact},
the determinant of $\mathcal{G}$ from \Cref{eqn:G_aS} can be computed as
\begin{align}
    \nonumber \det(\mathcal{G}) &= g_{\theta_{SA2}} (g_{\delta_3} g_{t_4} - g_{\delta_4} g_{t_3}) \\ \nonumber
    &= - \dfrac{ \cos(\gamma_S - \theta_{SA})}{R_{SA}} \Bigg[ \left( \frac{\cos(\gamma_D - \theta_{DA})}{R_{DA}} \right) \\ \nonumber
    &\quad \times \left( \frac{R_{DA} \cos(\gamma_D - \theta_{DA}) \mathcal{N}}{\mathcal{D}} \right) \\ \nonumber
    &\quad + \frac{\sin(\gamma_D - \theta_{DA})}{R_{DA}}\left( \frac{R_{DA} \sin(\gamma_D - \theta_{DA}) \mathcal{N}}{\mathcal{D}} \right)\Bigg] \\ \nonumber
    &= - \dfrac{ \cos(\gamma_S - \theta_{SA})}{R_{SA}}\frac{\mathcal{N}}{\mathcal{D}} \\
    &\quad \times \left[ \cos^2(\gamma_D - \theta_{DA}) + \sin^2(\gamma_D - \theta_{DA}) \right],\label{eqn:detG_aS}
\end{align}
where the terms $\mathcal{N}$ and $\mathcal{D}$ are defined as \begin{align*}
    \mathcal{N} &= (V_{R_{DA}} + 2\lambda)^2 - V_{\theta_{DA}}^2 + 2V_{\theta_{DA}}(V_{R_{DA}} + 2\lambda), \\
    \mathcal{D} &= (V_{R_{DA}}^2 + V_{\theta_{DA}}^2 + 2\lambda V_{R_{DA}})^2.
\end{align*}
\Cref{eqn:detG_aS} can be further simplified to obtain $\det(\mathcal{G})$ as
\begin{equation}
\begin{split}
    \det(\mathcal{G}) &= - \dfrac{ \cos(\gamma_S - \theta_{SA})}{R_{SA}} \\
    &\quad \times \left[ \frac{(V_{R_{DA}} + 2\lambda)^2 - V_{\theta_{DA}}^2 + 2V_{\theta_{DA}}(V_{R_{DA}} + 2\lambda)}{(V_{R_{DA}}^2 + V_{\theta_{DA}}^2 + 2\lambda V_{R_{DA}})^2} \right]. \label{eqn:detG_const_VS}
\end{split}
\end{equation}
There are two singularity points in \Cref{eqn:detG_const_VS}: first is when  $\gamma_S - \theta_{SA}=\pi/2$ and second is when $V_{\theta_{DA}} = 0$ and $V_{R_{DA}}=-2 \lambda$. 
It can be verified by computing the derivatives that $\gamma_S - \theta_{SA}=\pi/2$, if at all happen, will happen momentarily. The second case of singularity corresponds to a condition in which the defender and the attacker are on a collision course, with a closing speed of $2\lambda$. To avoid this case, the proportionality constant $\lambda$ can be chosen such that this condition never arises based on the initial engagement of the defender and the attacker. 
\subsection{Asset ship moving with variable speed ($\dot{V}_S \neq 0$) along a fixed direction ($a_S = 0$)}
In this subsection, we consider the scenario in which the asset ship cannot change direction but can actively control its speed. This reduces the control matrix $\mathcal{G}$ in \Cref{eqn:FG} to 
\begin{align}
    \mathcal{G} = \begin{bmatrix}
g_{\delta_1} & g_{\delta_3} & g_{\delta_4} \\
0 & g_{t_3} & g_{t_4}\\
g_{\theta_{SA1}} & 0 & 0
\end{bmatrix},\label{eqn:det_G_dVS}
\end{align}
whose determinant can be computed as $\det (\mathcal{G}) = g_{\theta_{SA1}} (g_{\delta_3}g_{t_4} - g_{\delta_4}g_{t_3})$. Using the expression of $g_{\theta_{SA1}}$, $g_{\delta_3}$, $g_{\delta_4}$, $g_{t_3}$, and $g_{t_4}$, as already derived (refer \Cref{eqn:theta_ddot_SA_compact,eqn:deltaddot_compact,eqn:tgodot_compact}), and noting that the  $(g_{\delta_3} g_{t_4} - g_{\delta_4} g_{t_3})$ was obtained for previous case, the determinant of $\mathcal{G}$  in \Cref{eqn:det_G_dVS} can be obtained as
\begin{equation}
\begin{split}
    \det(\mathcal{G}) &= - \dfrac{ \sin(\gamma_S - \theta_{SA}) }{R_{SA}} \\
    &\quad \times \left[ \frac{(V_{R_{DA}} + 2\lambda)^2 - V_{\theta_{DA}}^2 + 2V_{\theta_{DA}}(V_{R_{DA}} + 2\lambda)}{(V_{R_{DA}}^2 + V_{\theta_{DA}}^2 + 2\lambda V_{R_{DA}})^2} \right]. \label{eqn:detG_zero_aS}
\end{split}
\end{equation}
\Cref{eqn:detG_zero_aS} also has two singularity points: first is when $\gamma_S = \theta_{SA}$, and the second is same as previous case. As the asset ship cannot change its direction, $\gamma_S = \theta_{SA}$ corresponds to the case when the attacker approaches the asset ship along the line of motion of the asset ship. Note that, in this scenario, the asset ship can only change its speed to buy the defender more time to place itself on LOS, joining the asset ship and the attacker. 
\subsection{Defender is maneuvering ($a_D \neq 0$) with constant speed ($\dot{V}_D=0$)}
In this subsection, we consider a scenario in which the defender moves at a constant speed (depending on the attacker's speed) and steers its direction to capture the attacker. Thus, the control matrix in \Cref{eqn:FG} reduces to 
\begin{align}
    \mathcal{G} = \begin{bmatrix}
        g_{\delta_1} & g_{\delta_2} & g_{\delta_3}  \\
        0 & 0 & g_{t_3} \\
        g_{\theta_{SA1}} & g_{\theta_{SA2}} & 0
    \end{bmatrix},\label{eqn:G_aD}
\end{align}
whose determinant can be found out as $\det(\mathcal{G}) = g_{t_3} (g_{\delta_2} g_{\theta_{SA1}} - g_{\delta_1} g_{\theta_{SA2}})$. From the expressions of $g_\delta$ and $g_{\theta_{SA}}$ (refer \Cref{eqn:theta_ddot_SA_compact,eqn:deltaddot_compact,eqn:tgodot_compact}), it can be observed that $g_{\theta_{SA1}} = -g_{\delta_1}$ and $g_{\theta_{SA2}} = -g_{\delta_2}$. Thus, the determinant becomes zero, thereby making $\mathcal{G}$  in \Cref{eqn:G_aD}, a singular matrix. Thus, if the defender cannot change its speed, all of the objectives will not be met. However, following a similar approach, one can design another control strategy to achieve the interception of the attacker, with a reduced number of control objectives. Also, the constant speed case becomes similar to the standard interceptor guidance problem, that has been widely studied in references therein \cite{10.2514/6.2010-7876,10.1109/TAES.2025.3649483,10.1109/TCYB.2019.2958548,10.1109/LCSYS.2020.3041799}. Note from the expression of $\mathcal{G}$ matrix that the columns corresponding to $\dot{V}_S$ and $\dot{V}_D$ can be removed, so that it represents constant agent speed case. Subsequently, two rows relevant to the considered objectives can be used to design the control law.

\section{Simulation Results}\label{sec:simulation}
In this section, we demonstrate the efficacy of the proposed strategy through numerical simulations in MATLAB\textsuperscript{\textregistered}. 
We tested the proposed control law as given by \Cref{eqn:U}, for various initial configurations of the attacker and the defender. 
Furthermore, the proposed cooperative control strategy for joint maneuver of the asset ship and the defender is also tested against an attacker utilizing various guidance laws, including proportional navigation (PN), realistic true proportional navigation (RTPN), and augmented proportional navigation (APN).
In addition to the aforementioned validations, we also demonstrate the working of the proposed strategy across varying levels of control authority available to the asset ship and the defender. 
The controller gain parameters chosen for the simulations are as follows: $\mathcal{M}_1 = 0.1$, $\mathcal{M}_2 = 0.1$, $\mathcal{M}_3 = 0.02$, and $k_\delta = 10$. The proportionality constant for attacker guidance law, $\lambda$, is $20$.  
The input acceleration bounds (in m/s$^2$) for the asset are $|\dot{V}_S|\leq 0.1$, $|a_S|\leq 0.1$; for the defender are $|\dot{V}_D|\leq 10$, $|a_D|\leq 10$; and lastly for the attacker are
$|\dot{V}_A|\leq 10$, $|a_A|\leq 10$.

\begin{figure*}[!ht]
    \centering
    \begin{subfigure}{0.33\linewidth}
    \centering
    \includegraphics[width=\linewidth]{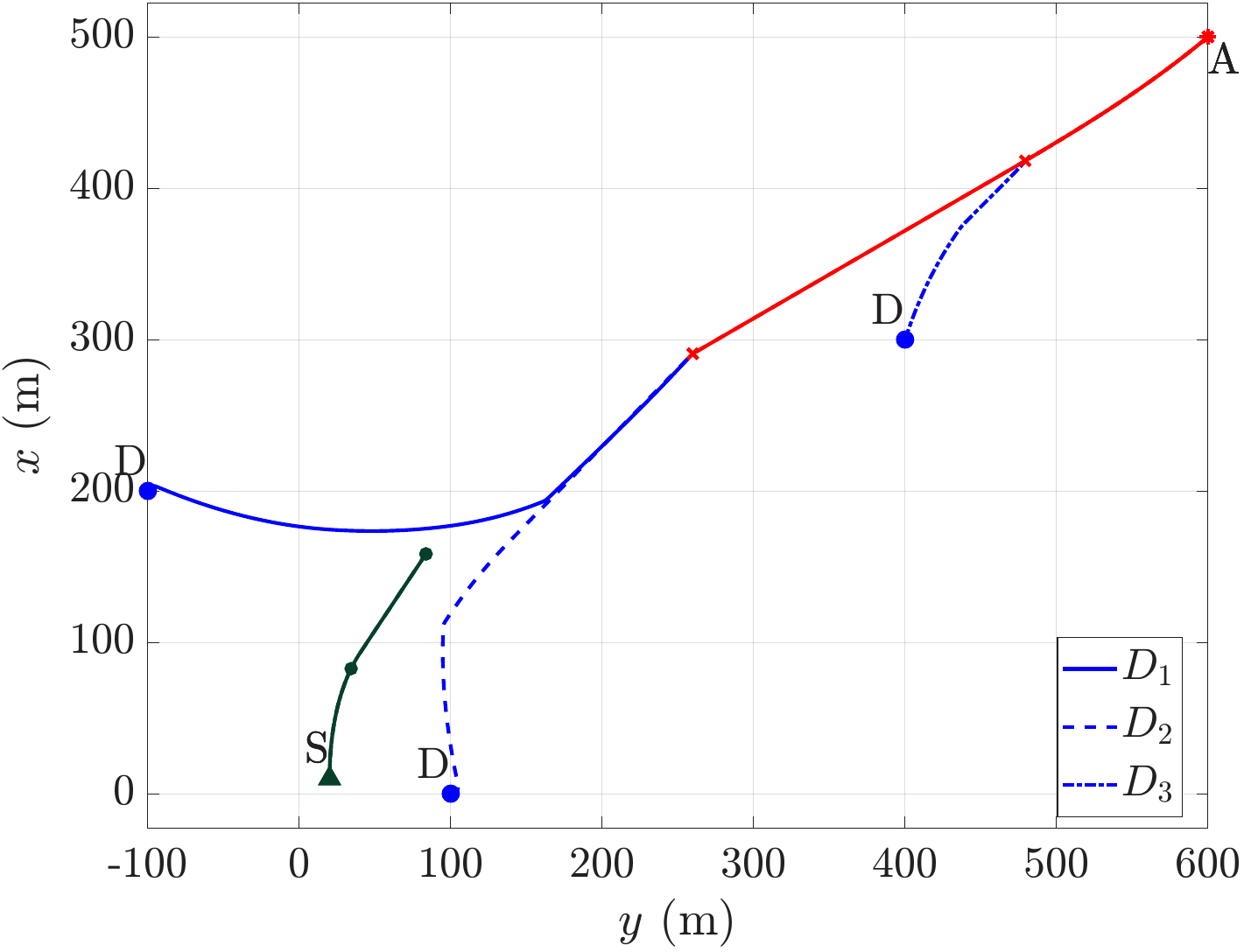}   
    \caption{Trajectories of asset, defender, and attacker.}
    \label{fig:D_path_indp_aD_dvD}
    \end{subfigure}%
    \begin{subfigure}{0.33\linewidth}
    \centering
     \includegraphics[width=\linewidth]{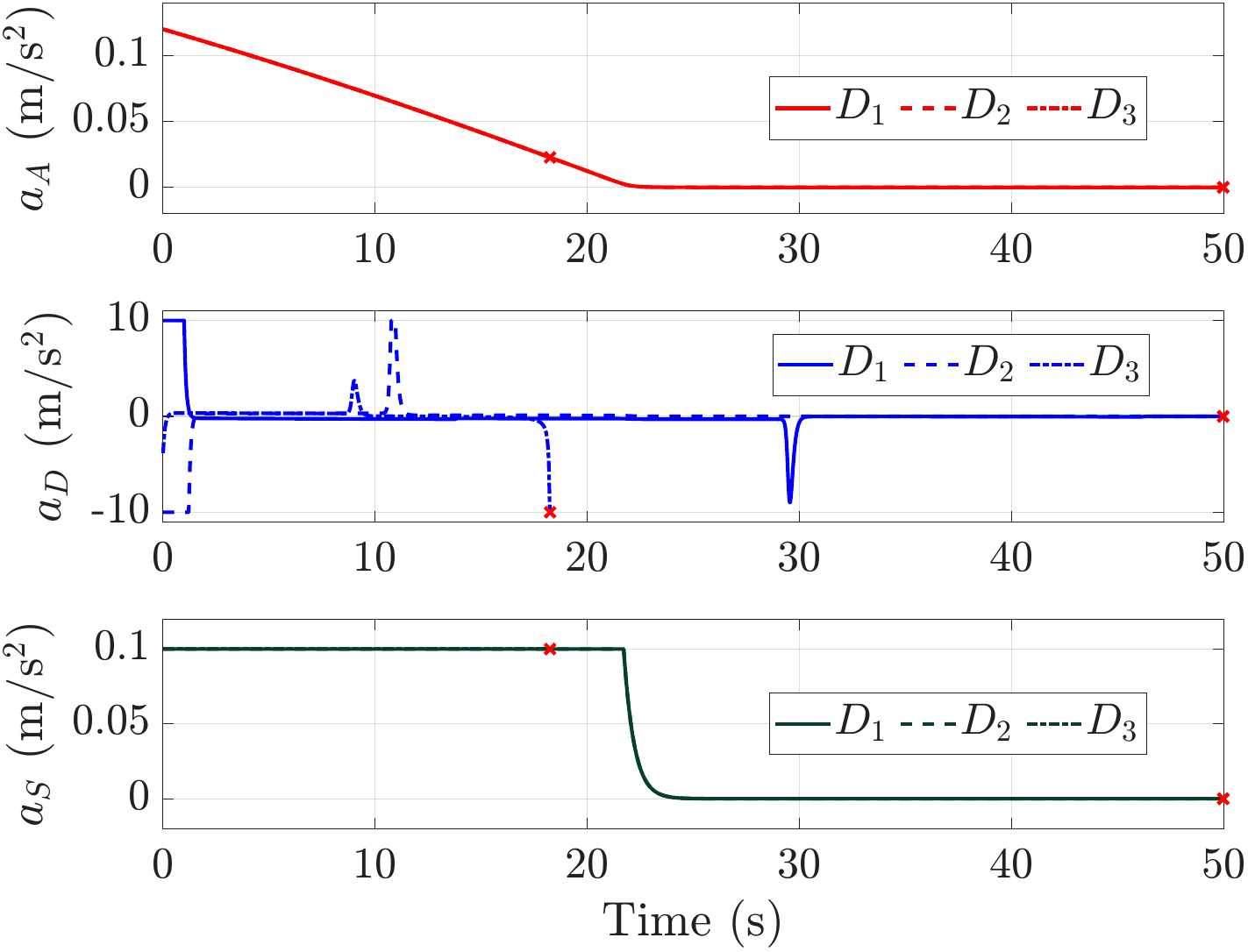}
    \caption{Control inputs: lateral acceleration.}
    \label{fig:D_latax_indp_aD_dvD}
    \end{subfigure}%
    \begin{subfigure}{0.33\linewidth}
    \centering
     \includegraphics[width=\linewidth]{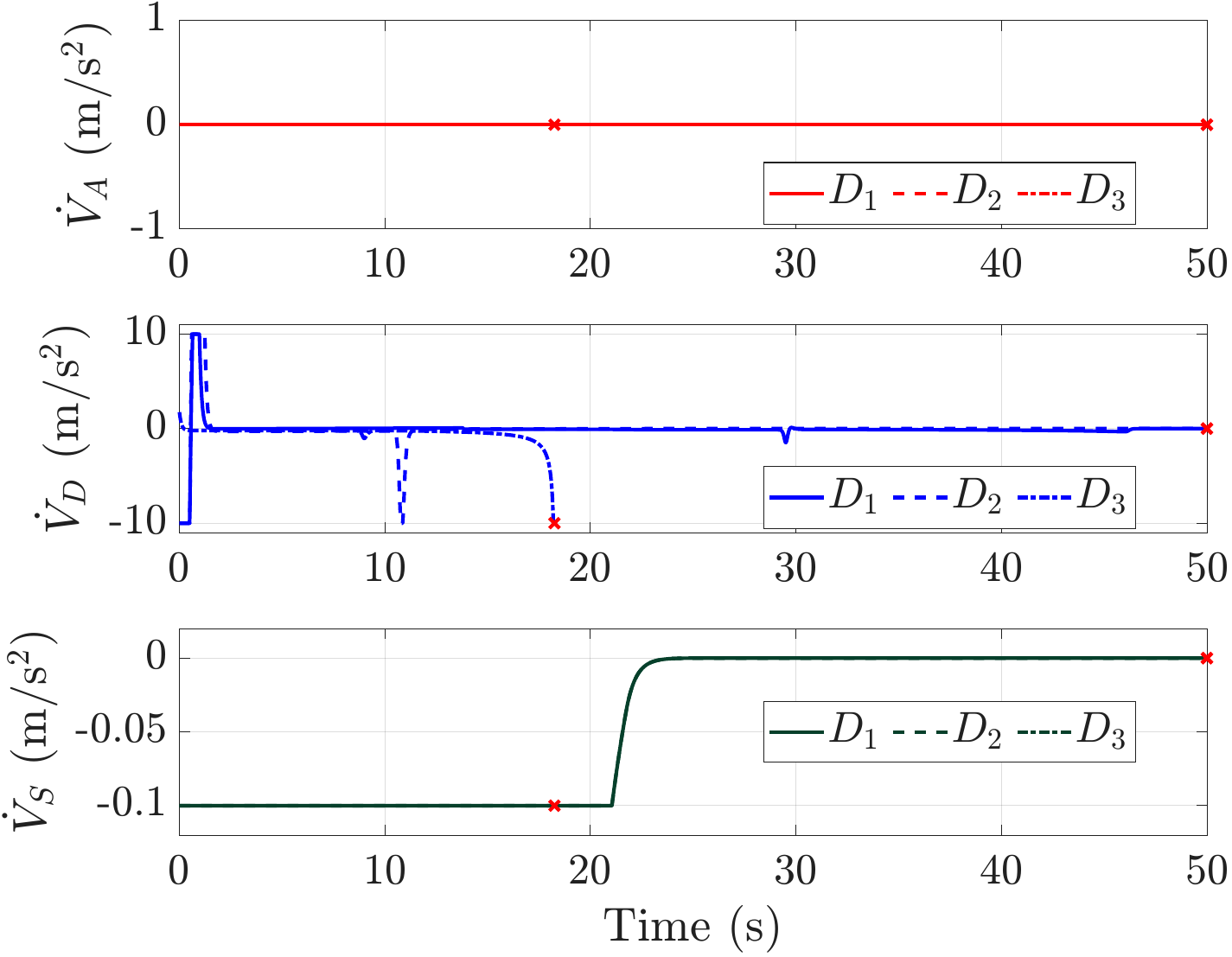}
    \caption{Control inputs: linear acceleration.}
    \label{fig:D_linAcc_indp_aD_dvD}
    \end{subfigure}
    \begin{subfigure}{0.33\linewidth}
    \centering
    \includegraphics[width=\linewidth]{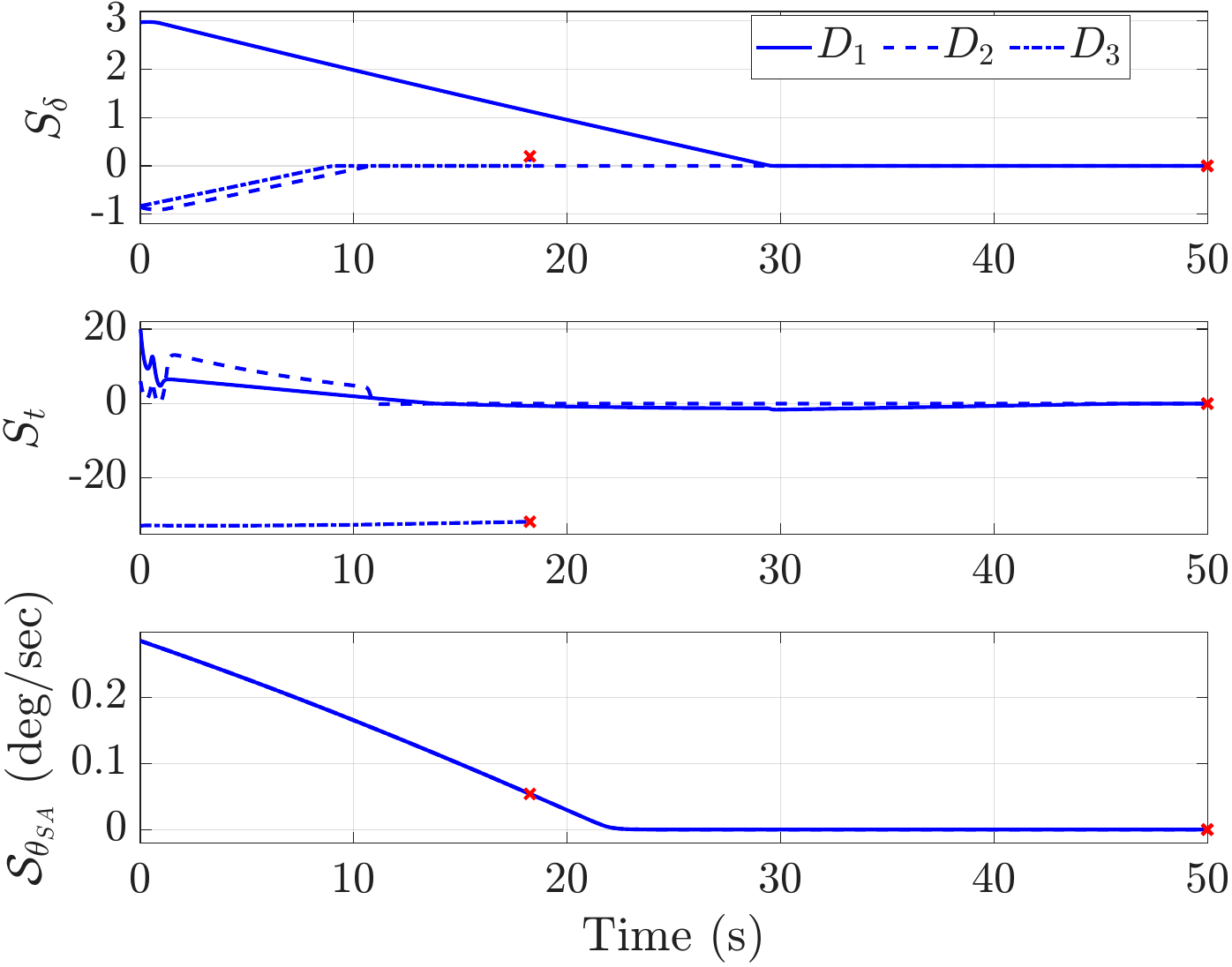}
    \caption{Sliding surfaces.}
    \label{fig:D_surface_indp_aD_dvD}
    \end{subfigure}%
    \begin{subfigure}{0.33\linewidth}
    \centering
    \includegraphics[width=\linewidth]{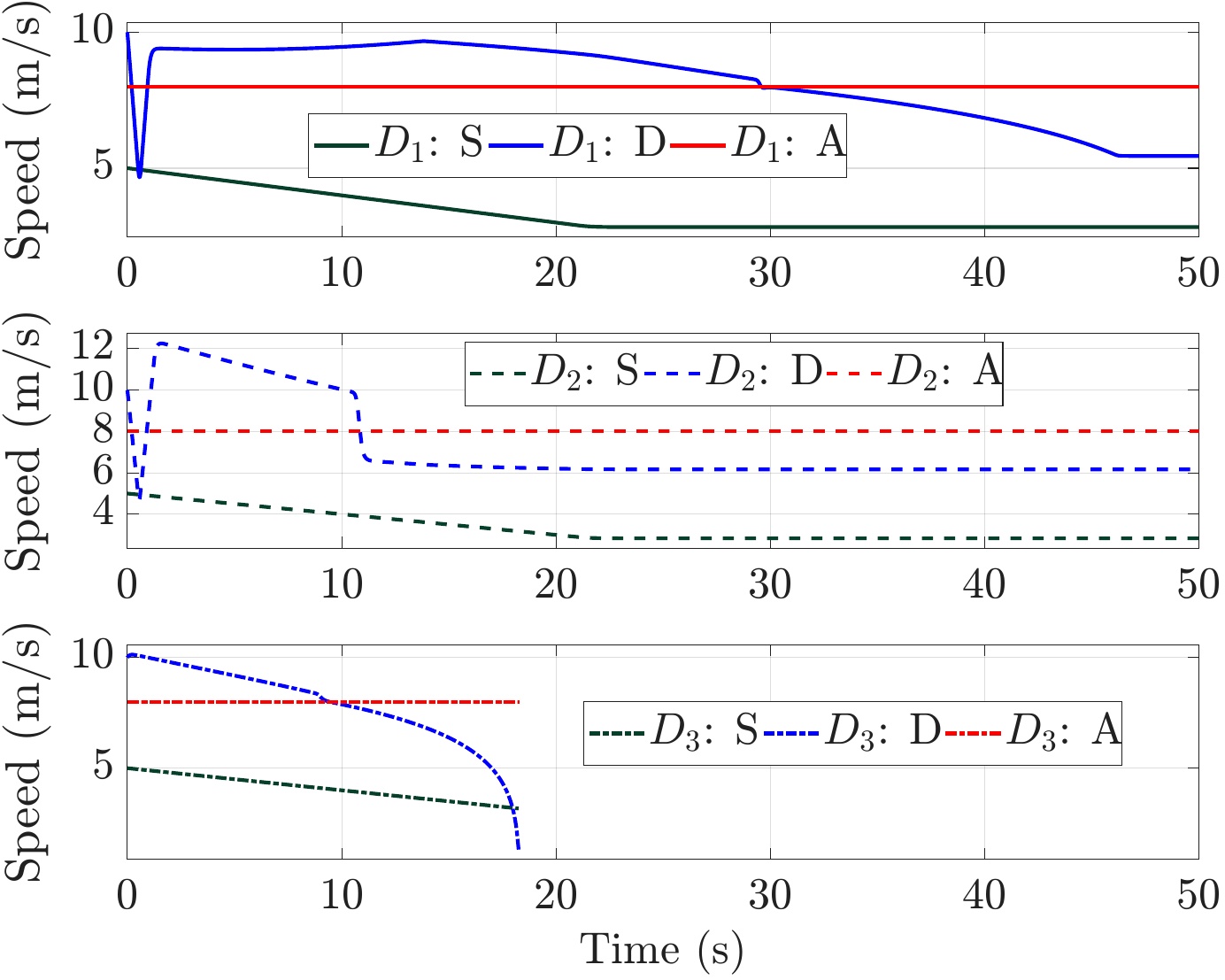}
    \caption{Comparison of speeds of the agents.}
    \label{fig:D_Speed_indp_aD_dvD}
    \end{subfigure}%
    \begin{subfigure}{0.33\linewidth}
    \centering
    \includegraphics[width=\linewidth]{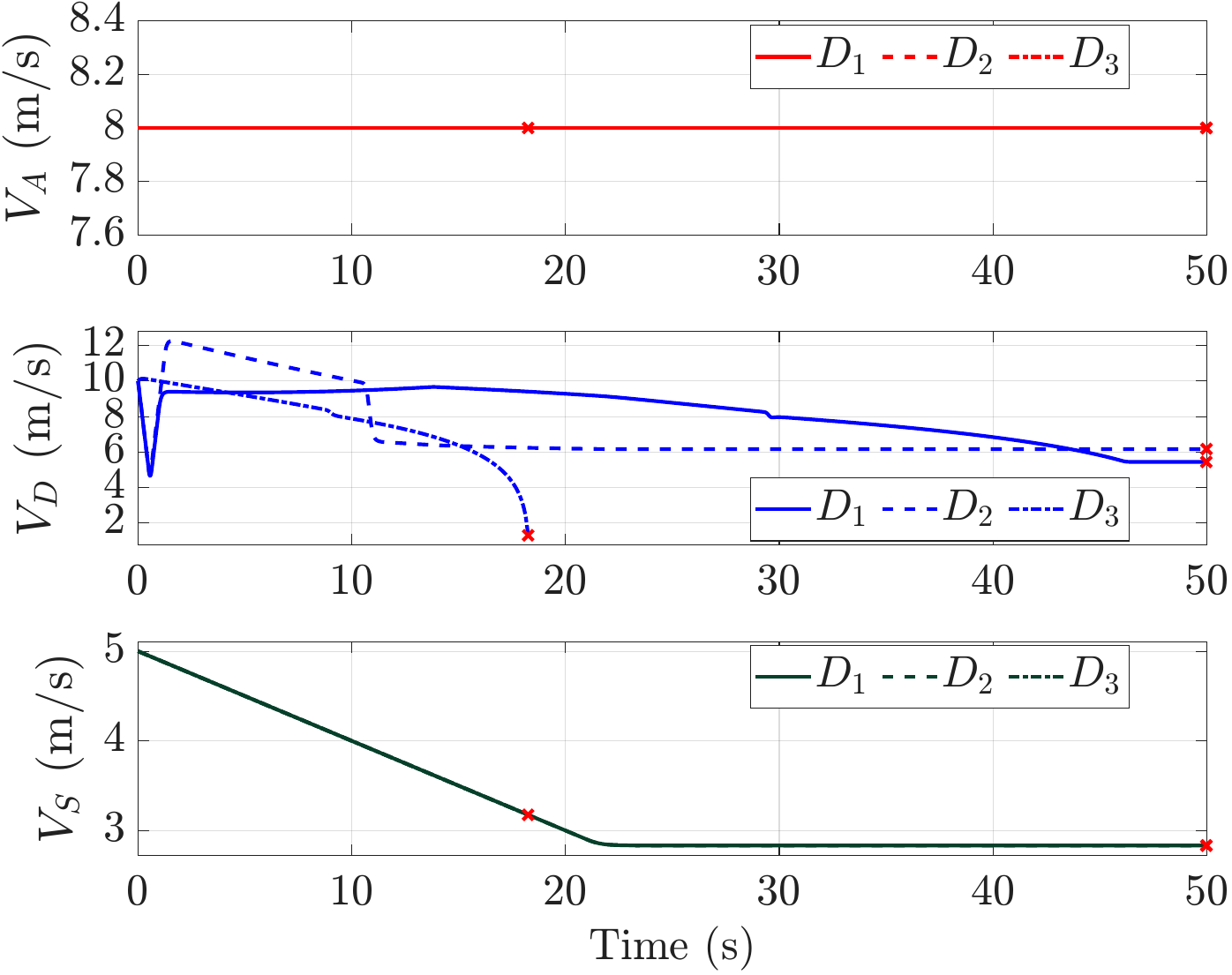}
    \caption{Speed variation of the agents.}
    \label{fig:D_speedAgentwise_indp_aD_dvD}
    \end{subfigure}
    \caption{Interception of the attacker for different initial locations of the defender.}
    \label{figcase:PN_defender}
\end{figure*}

First, we validate the proposed strategy for three different initial configurations of the defender, denoted by $D_1$, $D_2$, and $D_3$, in the subsequent discussions. The corresponding simulation results are presented in \Cref{figcase:PN_defender}.
In this simulation scenario, the asset ship ($S$) is initially placed at $10$ m North and $20$ m East and is moving towards North at a speed of $5$ m/s. 
The attacker is coming from $500$ m North and $ 600$ m East, with an initial course of $-130^\circ$ from North and approaching at a speed of $8$ m/s.
In case $D_1$, the defender ($D$) is located at $200$ m North and $100$ m West, heading towards $20^\circ$ from North at $10$ m/s. 
In the second case, $D_2$, the defender is at $100$ m East, moving East at $10$ m/s. 
In the last case, $D_3$, the defender is moving at $10$ m/s with a heading of $20^\circ$ from the North, starting from the position coordinates $(300,\, 400)$ m.
For these engagement scenarios of the cases $D_1$, $D_2$, and $D_3$, the desired time, $T_d$, at which the defender should capture the attacker is set to $50$ seconds. 
The paths taken by the asset, the defender, and the attacker are depicted in \Cref{fig:D_path_indp_aD_dvD}. It can be seen that the defender successfully captures the attacker before it could reach the asset. The terminal point of the engagement is denoted by a cross mark ($\times$) in all the simulation figures, unless stated otherwise. 
The steering commands issued to the asset and the defender are shown in \Cref{fig:D_latax_indp_aD_dvD,fig:D_linAcc_indp_aD_dvD}. 
It can be observed that the asset continuously steers itself to nullify the LOS rate between itself and the attacker, as per its control capability. Because the asset is having very little maneuverability, the commanded lateral acceleration ($a_S$) saturates (refer \Cref{fig:D_latax_indp_aD_dvD}).
The asset also decelerates to lower its speed. The defender also steers its position to be on the asset-attacker LOS and controls its speed to intercept the incoming attacker at the designated time. The temporary increase of $a_D$ in \Cref{fig:D_latax_indp_aD_dvD}, followed by drop in speed in \Cref{fig:D_Speed_indp_aD_dvD,fig:D_speedAgentwise_indp_aD_dvD}, correctly depicts the expected behavior as the defender slows down while making sharper turn.
From the $a_A$ profile as shown in \Cref{fig:D_latax_indp_aD_dvD}, it can be observed that the attacker maneuver is reduced. This is because the asset is continuously maneuvering to put the attacker on a collision course. The speed variation of each of the agents is also shown in \Cref{fig:D_Speed_indp_aD_dvD}.
From \Cref{fig:D_surface_indp_aD_dvD}, it is evident that the asset ship and the defender could jointly maneuver to meet all three objectives, as the sliding surfaces are converging towards zero.
The convergence of $\mathcal{S}_\delta$ to zero indicates that the defender could position itself on the LOS between the asset and the attacker. 
The convergence of $\mathcal{S}_t$ to zero indicates that the time constraints are satisfied. Lastly, $\mathcal{S}_{\theta_{SA}}$ going to zero means that the LOS between the asset ship and the attacker became non-rotating and the two are on a collision course. However, for the case of $D_3$, only $\mathcal{S}_\delta$ went to zero. It is to be noted that, in this case, the defender has been placed closer to the attacker's location. Therefore, it is worth highlighting that the defender could position itself on the LOS between the asset and the attacker, and the attacker, on its way to the asset, meets its end. Towards meeting the time constraint, the defender also seems to reduce its speed (refer \Cref{fig:D_Speed_indp_aD_dvD}). 

\begin{figure*}[!ht]
    \centering
    \begin{subfigure}{0.33\linewidth}
    \centering
    \includegraphics[width=\linewidth]{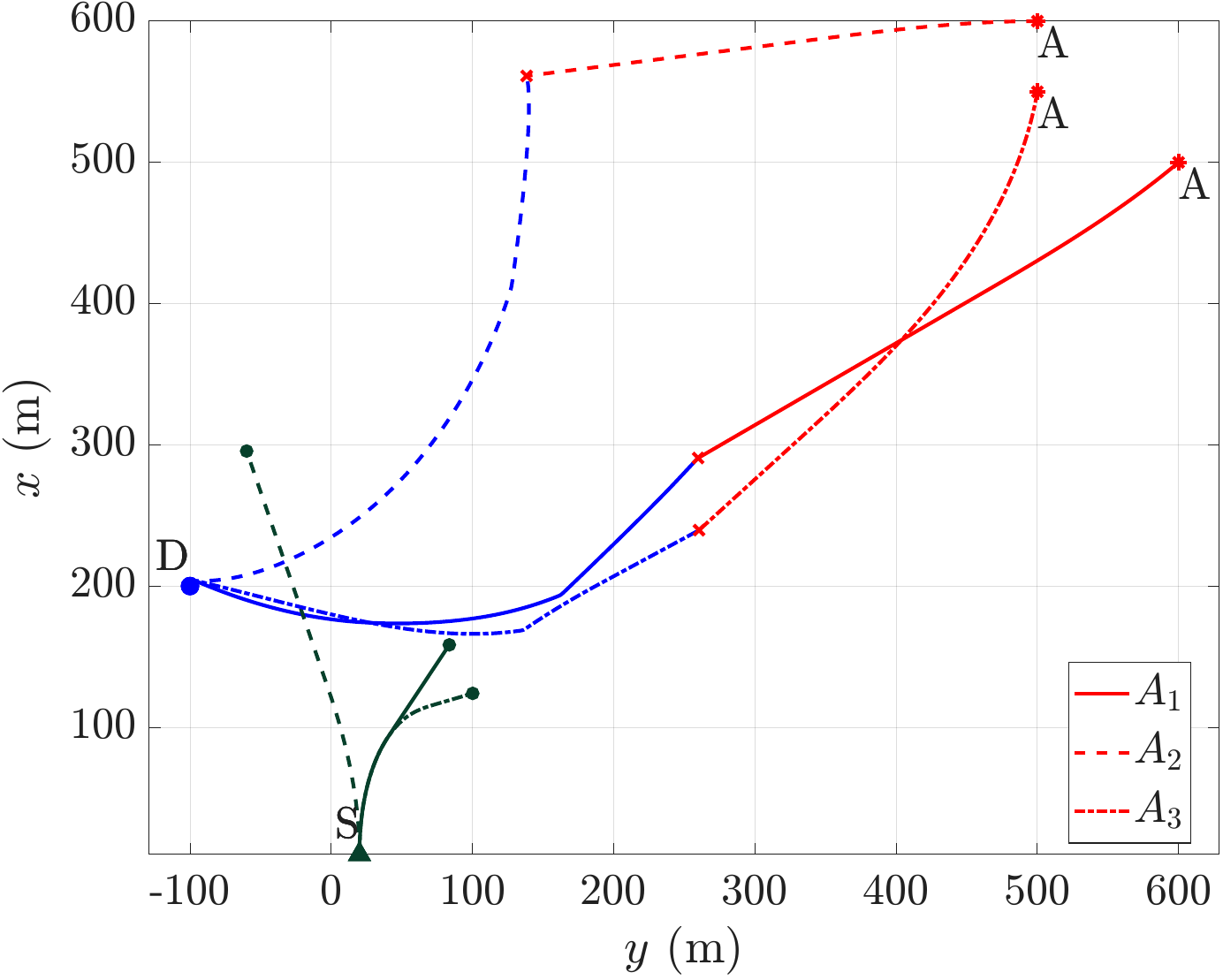}   
    \caption{Trajectories of asset, defender, and attacker.}
    \label{fig:A_path_indp_aD_dvD}
    \end{subfigure}%
    \begin{subfigure}{0.33\linewidth}
    \centering
     \includegraphics[width=\linewidth]{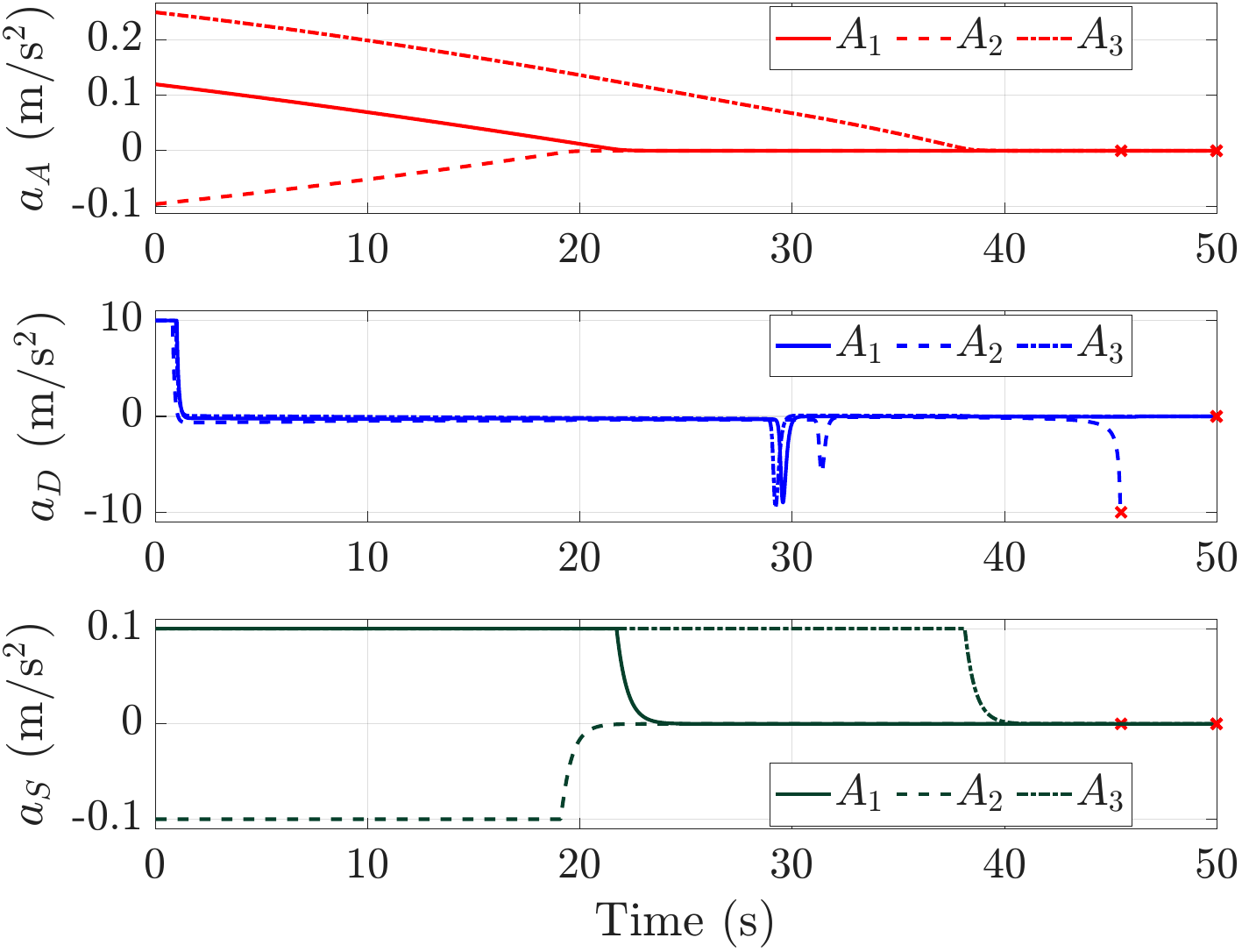}
    \caption{Control inputs: lateral acceleration.}
    \label{fig:A_latax_indp_aD_dvD}
    \end{subfigure}%
    \begin{subfigure}{0.33\linewidth}
    \centering
     \includegraphics[width=\linewidth]{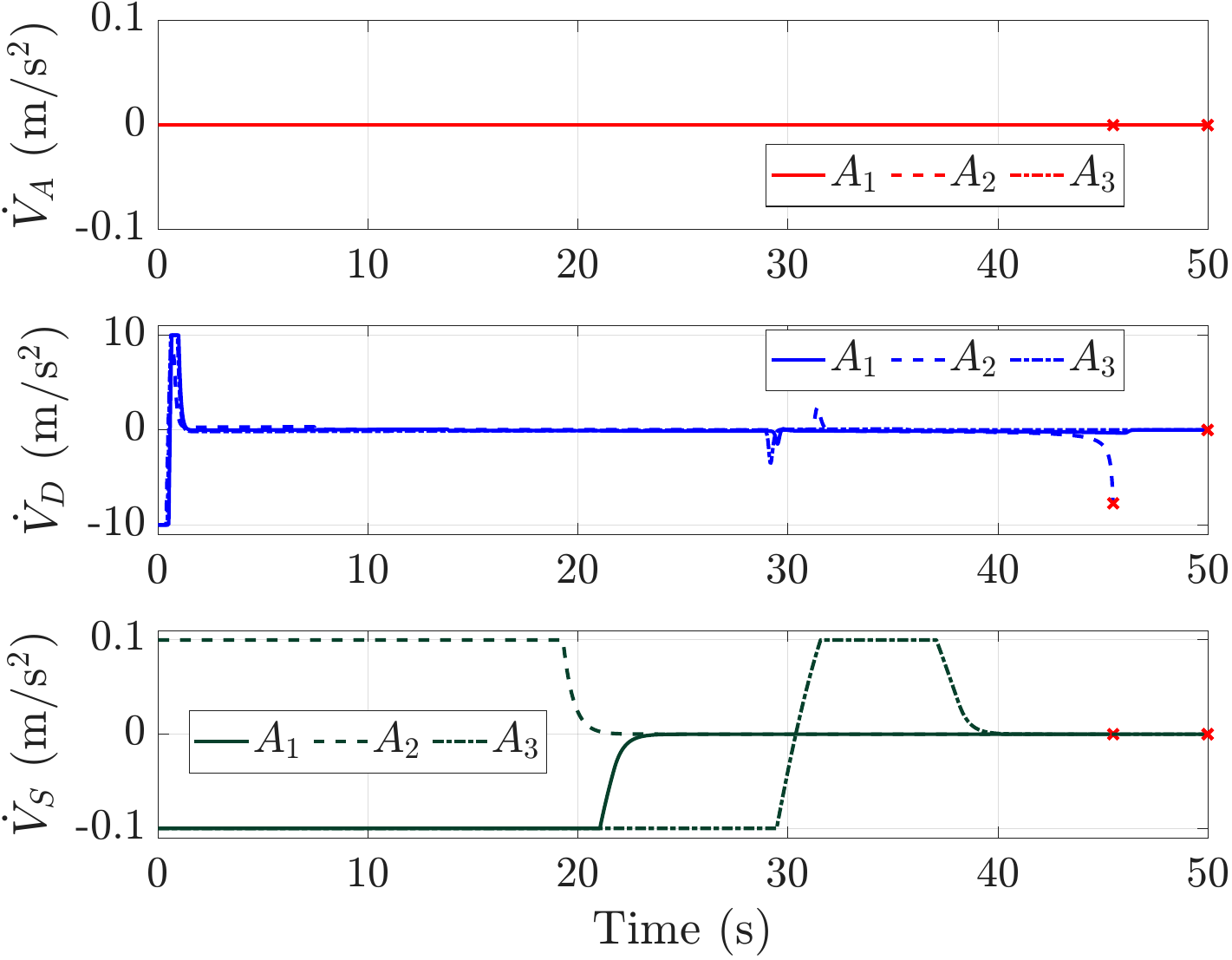}
    \caption{Control inputs: linear acceleration.}
    \label{fig:A_linAcc_indp_aD_dvD}
    \end{subfigure}
    \begin{subfigure}{0.33\linewidth}
    \centering
    \includegraphics[width=\linewidth]{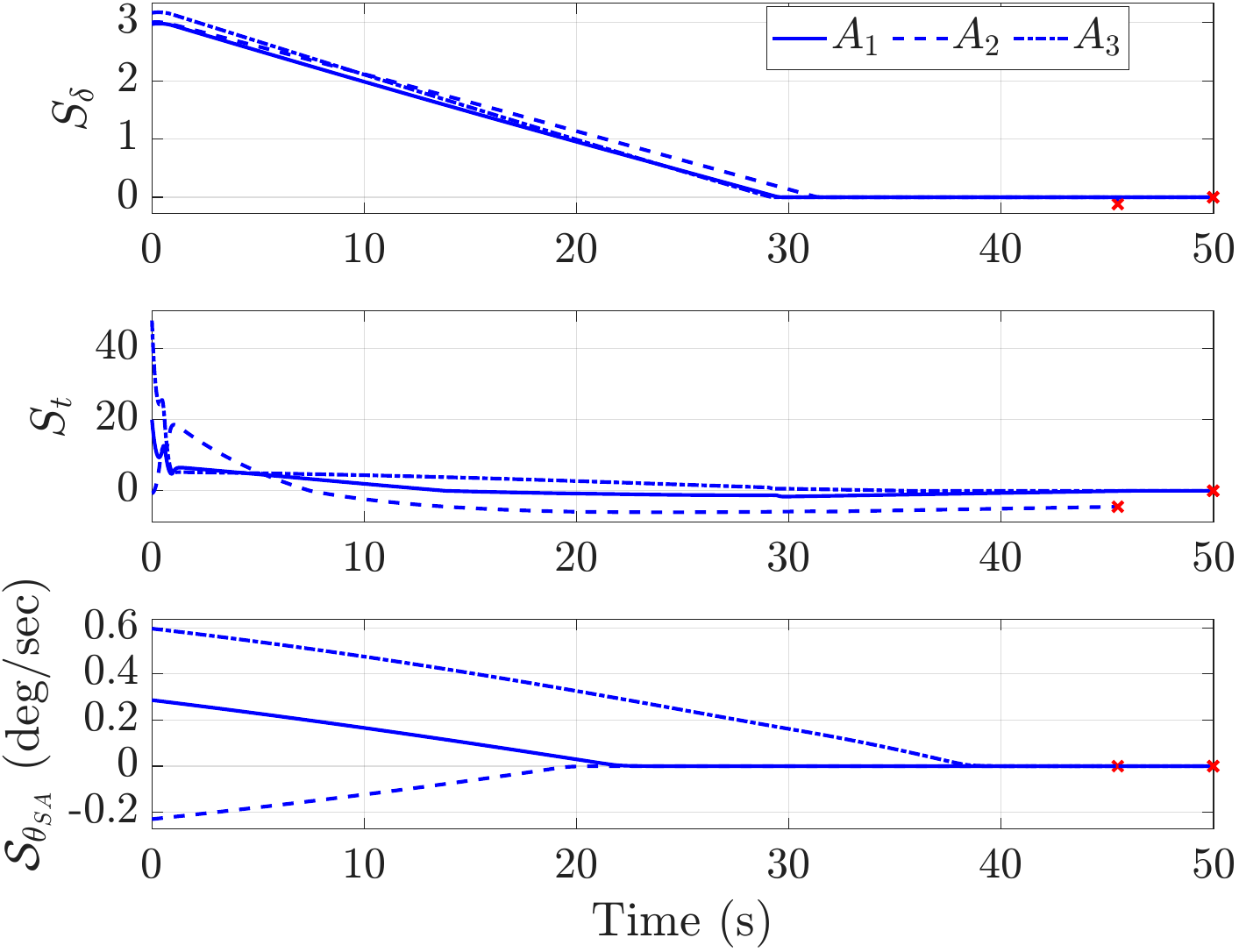}
    \caption{Sliding surfaces.}
    \label{fig:A_surface_indp_aD_dvD}
    \end{subfigure}%
    \begin{subfigure}{0.33\linewidth}
    \centering
    \includegraphics[width=\linewidth]{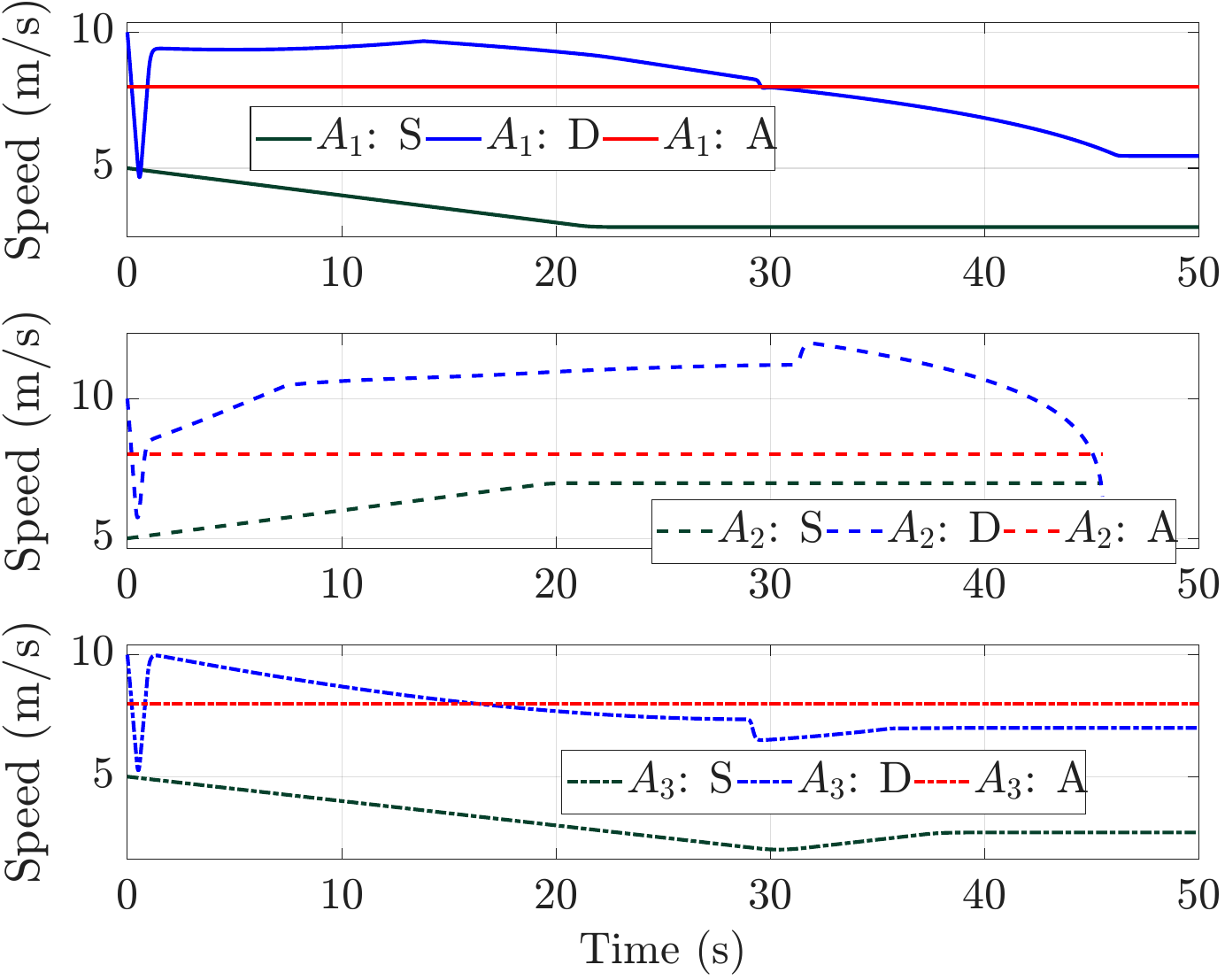}
    \caption{Comparison of speeds of the agents.}
    \label{fig:A_Speed_indp_aD_dvD}
    \end{subfigure}%
    \begin{subfigure}{0.33\linewidth}
    \centering
    \includegraphics[width=\linewidth]{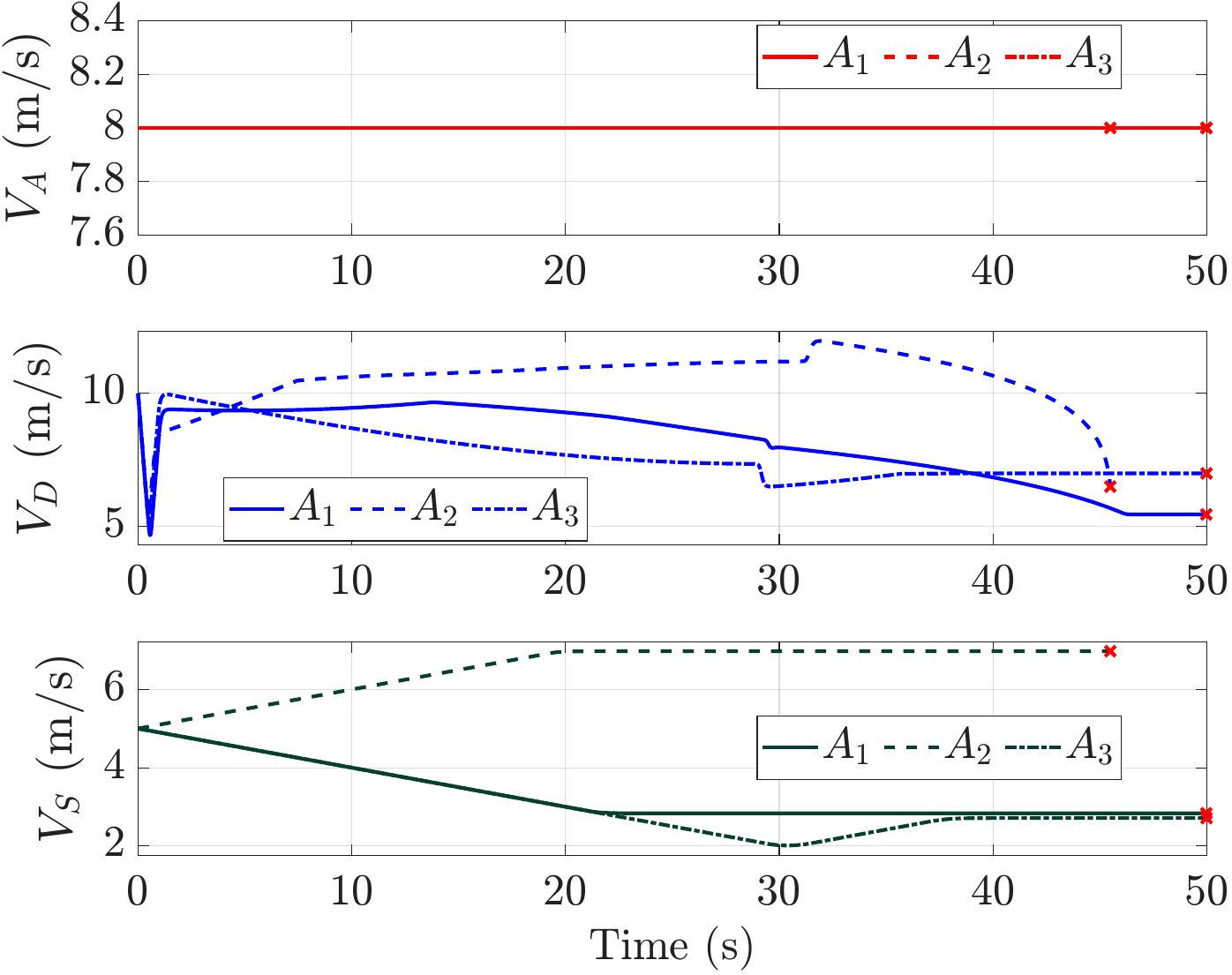}
    \caption{Speed variation of the agents.}
    \label{fig:A_speedAgentwise_indp_aD_dvD}
    \end{subfigure}
    \caption{Interception of the attacker for different initial locations of the attacker.}
    \label{figcase:PN_attacker}
\end{figure*}

Next, we validate the control strategy for various initial attacker locations (see \Cref{figcase:PN_attacker}). The states of the agents involved are represented by a tuple $(x,\, y,\, V,\, \gamma)$ where $x$ is the North coordinate, $y$ is the East coordinate, $V$ is the speed of the agent, and $\gamma$ is the direction from the North.
The initial configuration of the asset ship is $S:~(10\text{~m},~20\text{~m},~5\text{~m/s},~0^\circ)$, the defender is $D:~(200\text{~m},~-100\text{~m},~10\text{~m/s},~20^\circ)$.
The three initial configurations of the attacker under consideration are taken as $A_1:~(500\text{~m},~600\text{~m},~8\text{~m/s},~-130^\circ)$, $A_2:~(600\text{~m},~500\text{~m},~8\text{~m/s},~-90^\circ)$, and $A_3:~(550\text{~m},~500\text{~m},~8\text{~m/s},~-170^\circ)$.
The paths of the three agents are depicted in \Cref{fig:A_path_indp_aD_dvD}, illustrating a successful interception of the attacker by the defender for all three cases. The corresponding control inputs (lateral accelerations and linear accelerations) are shown in \Cref{fig:A_latax_indp_aD_dvD,fig:A_linAcc_indp_aD_dvD}.
The sliding surfaces (refer \Cref{fig:A_surface_indp_aD_dvD}) go to zero, indicating that the joint maneuver of the defender and the attacker could achieve the set objectives. However, in the case of $A_2$, the time constraint is not satisfied, but as desired the attacker intercepts the defender before it could reach the asset ship. 
The speeds of each agent are also shown in \Cref{fig:A_Speed_indp_aD_dvD,fig:A_speedAgentwise_indp_aD_dvD}. From the defender's speed profile in case $A_2$, it can be seen that it reduces its speed to meet the time constraint. 

In all the previous cases, the attacker followed a PN guidance law. However, it is necessary to validate the proposed strategy against various guidance laws commonly employed in adversarial situations. 
\begin{figure*}[!ht]
    \centering
    \begin{subfigure}{0.33\linewidth}
    \centering
    \includegraphics[width=\linewidth]{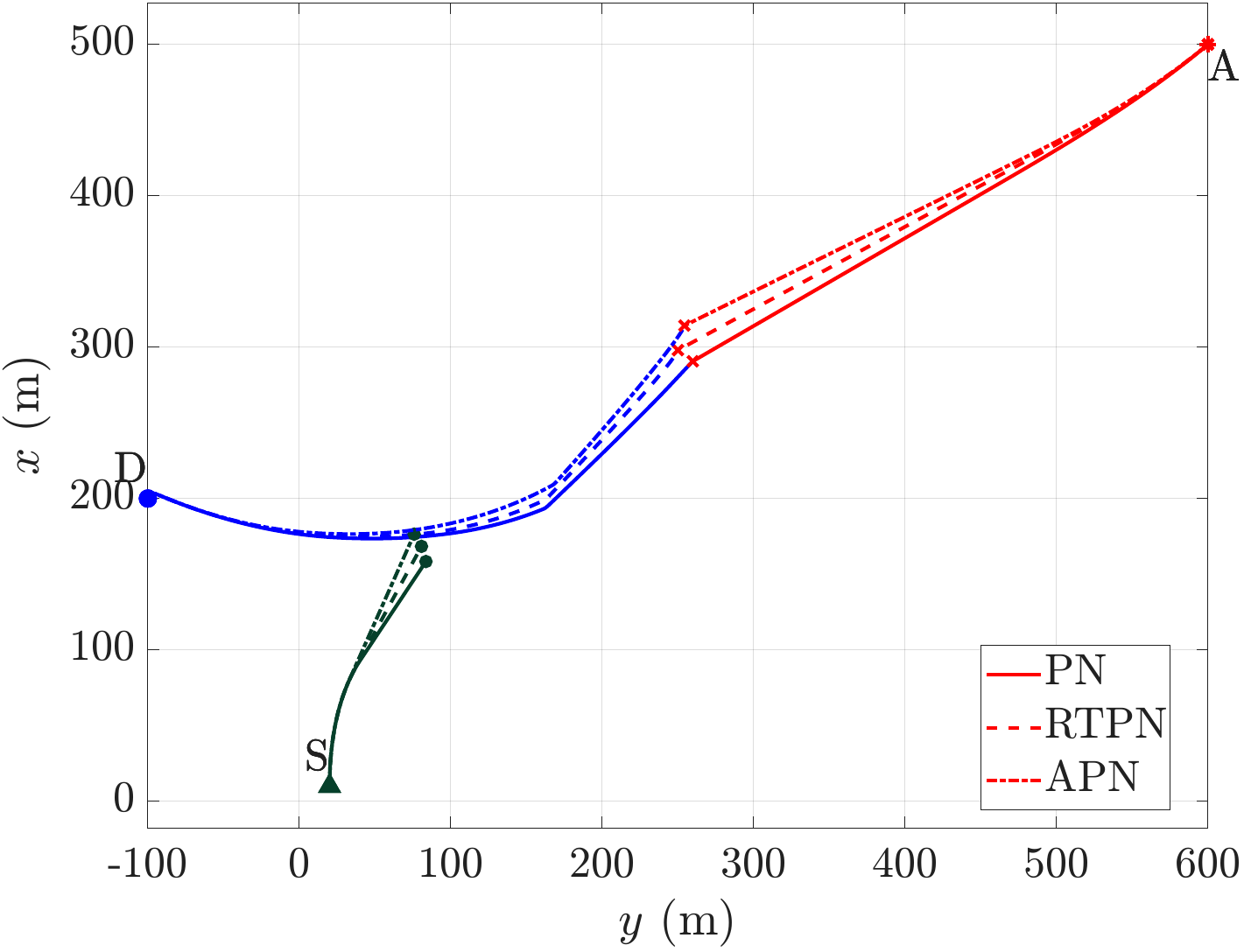}   
    \caption{Trajectories of asset, defender, and attacker.}
    \label{fig:L_path_indp_aD_dvD}
    \end{subfigure}%
    \begin{subfigure}{0.33\linewidth}
    \centering
     \includegraphics[width=\linewidth]{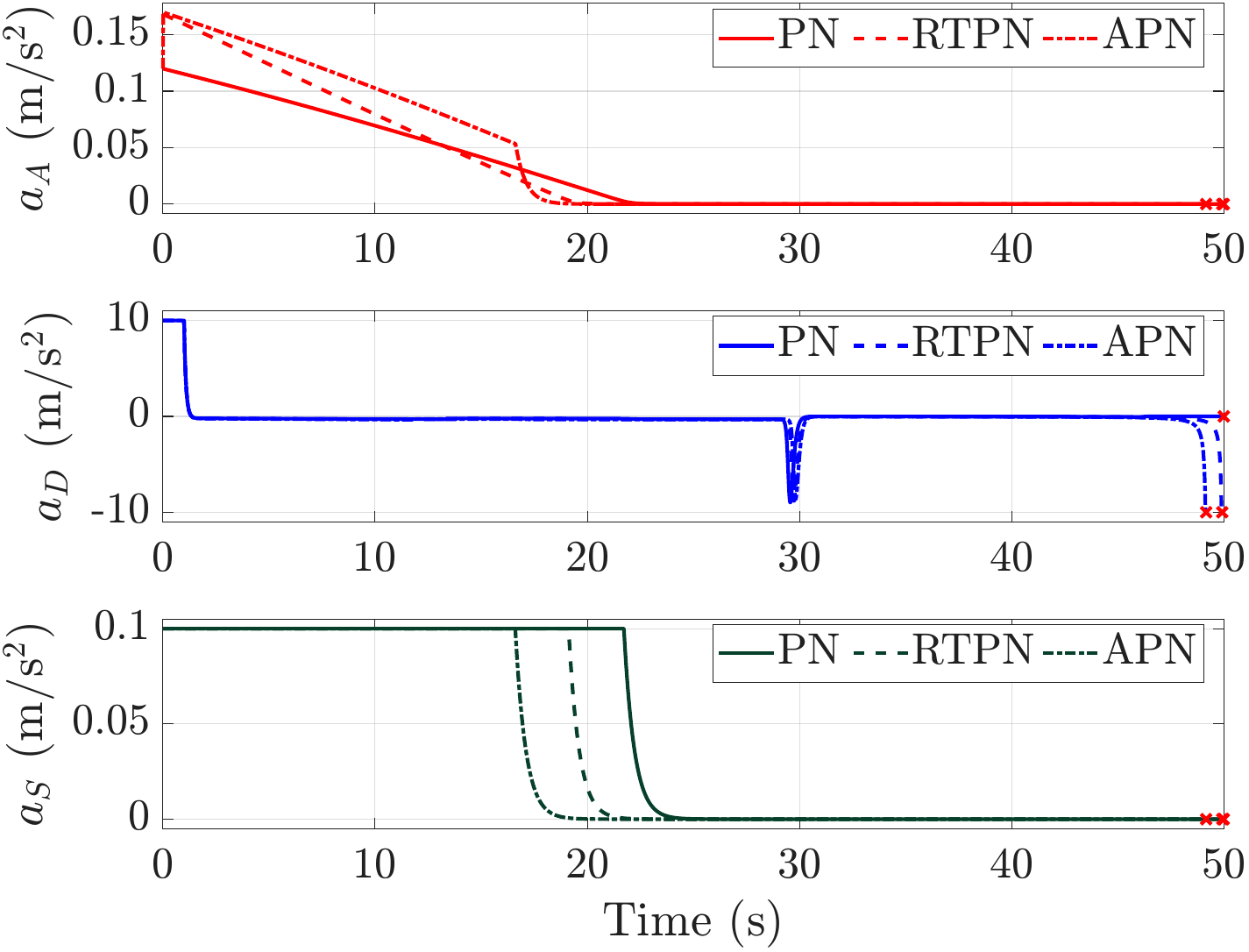}
    \caption{Control inputs: lateral acceleration.}
    \label{fig:L_latax_indp_aD_dvD}
    \end{subfigure}%
    \begin{subfigure}{0.33\linewidth}
    \centering
     \includegraphics[width=\linewidth]{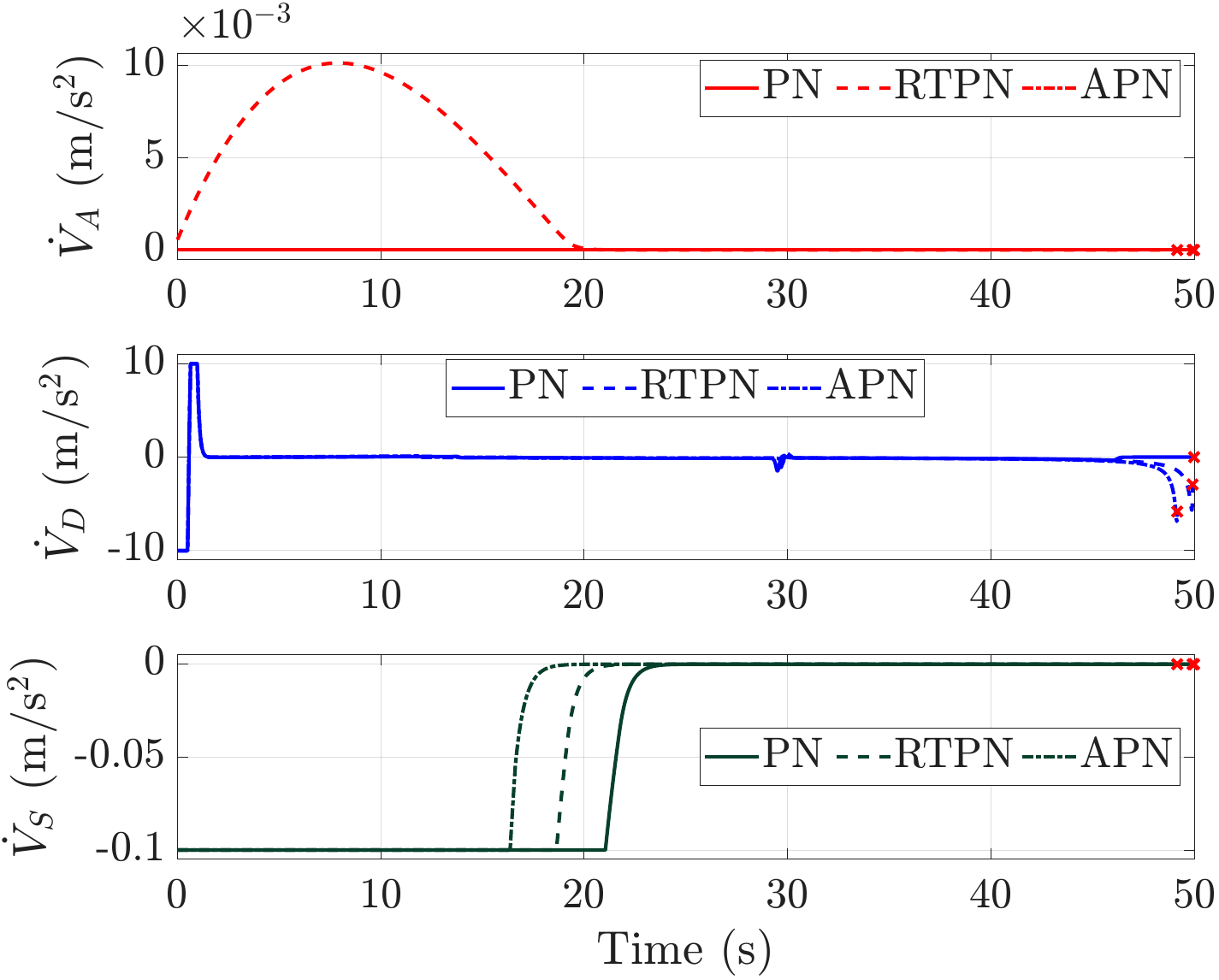}
    \caption{Control inputs: linear acceleration.}
    \label{fig:L_linAcc_indp_aD_dvD}
    \end{subfigure}
    \begin{subfigure}{0.33\linewidth}
    \centering
    \includegraphics[width=\linewidth]{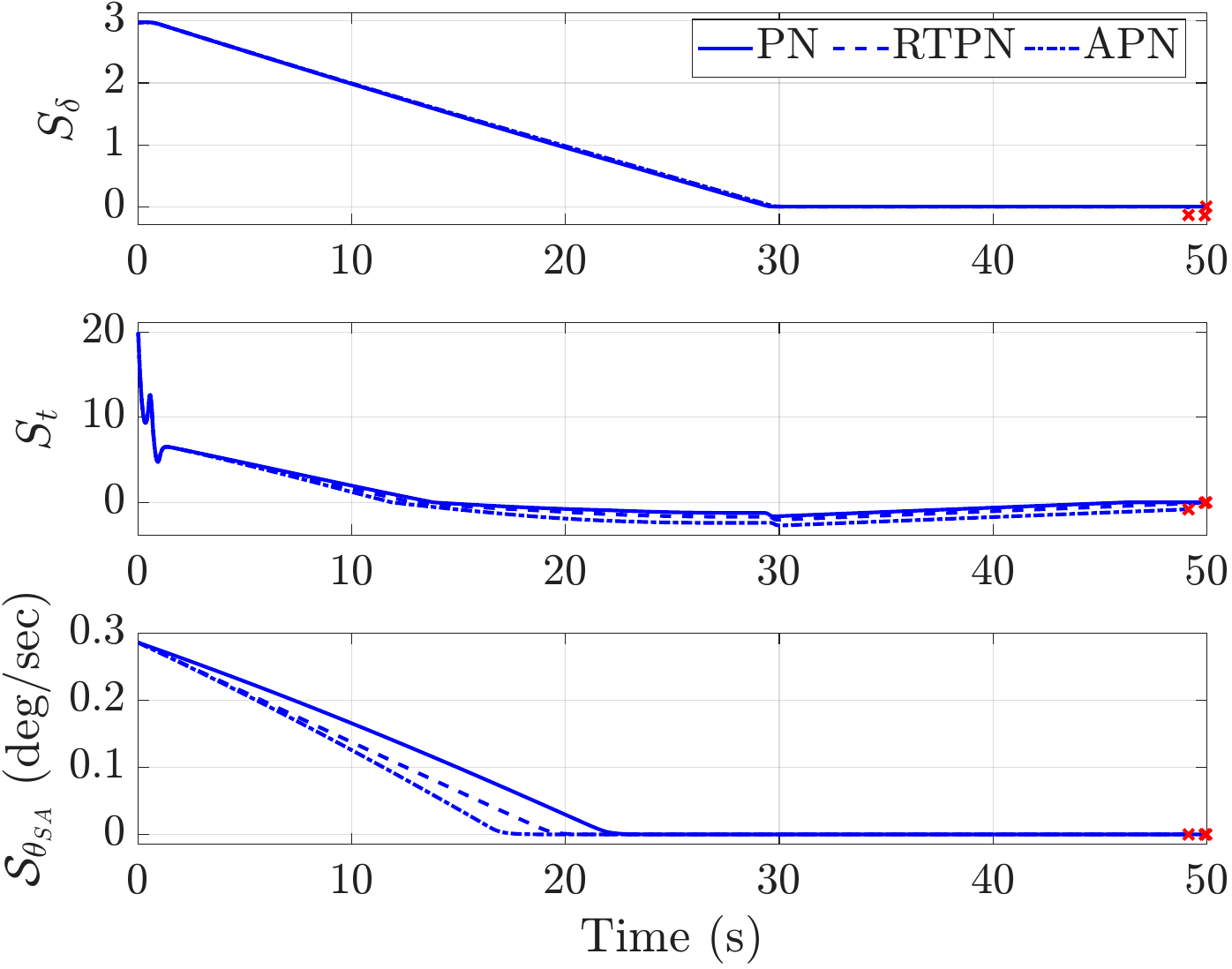}
    \caption{Sliding surfaces.}
    \label{fig:L_surface_indp_aD_dvD}
    \end{subfigure}%
    \begin{subfigure}{0.33\linewidth}
    \centering
    \includegraphics[width=\linewidth]{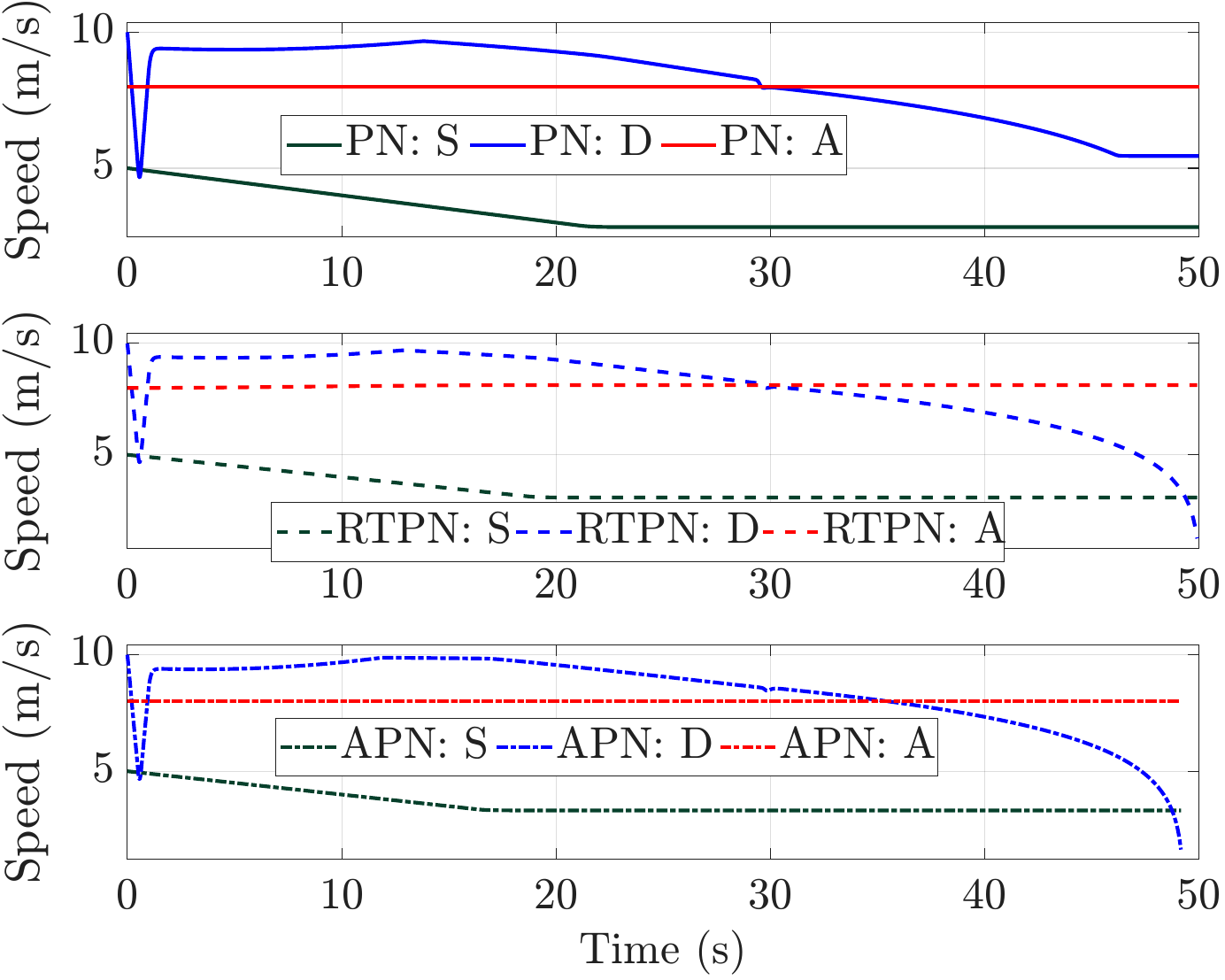}
    \caption{Comparison of speeds of the agents.}
    \label{fig:L_Speed_indp_aD_dvD}
    \end{subfigure}%
    \begin{subfigure}{0.33\linewidth}
    \centering
    \includegraphics[width=\linewidth]{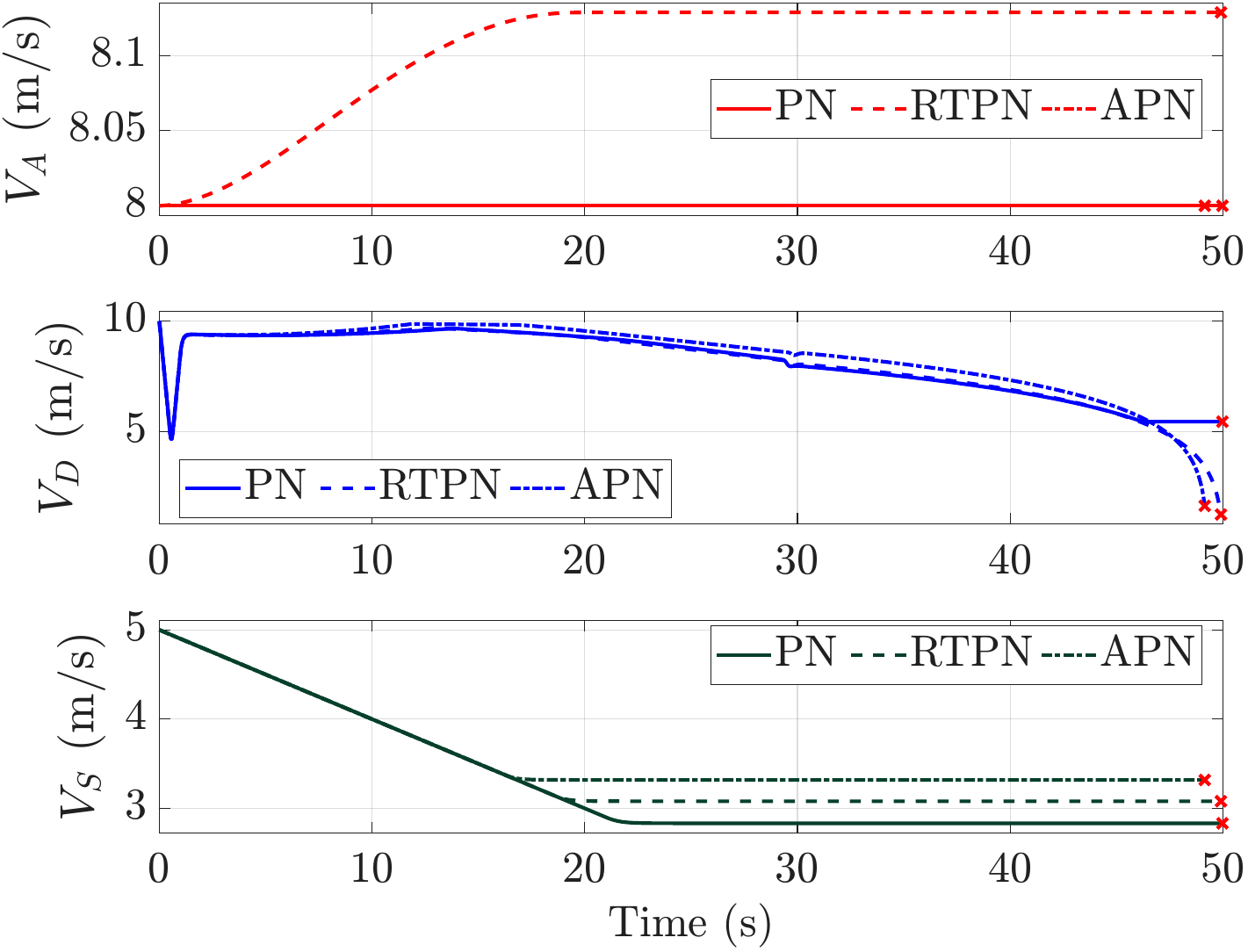}
    \caption{Speed variation of the agents.}
    \label{fig:L_speedAgentwise_indp_aD_dvD}
    \end{subfigure}
    \caption{Interception of the attacker when the attacker pursues various guidance laws.}
    \label{figcase:L_attackerlaws}
\end{figure*}
Therefore, we consider three of the most commonly employed guidance laws: PN, RTPN, and APN. The effectiveness of the proposed strategy for capturing attackers using various guidance laws is depicted in \Cref{figcase:L_attackerlaws}. 
The initial configurations of the agents ($S,~D,~A$) taken for the simulation are  $S:~(10\text{~m},~20\text{~m},~5\text{~m/s},~0^\circ)$, $D:~(200\text{~m},~-100\text{~m},~10\text{~m/s},~20^\circ)$, and  $A:~(500\text{~m},~600\text{~m},~8\text{~m/s},~-130^\circ)$. 
The paths depicted in \Cref{fig:L_path_indp_aD_dvD} show a successful interception of the attacker irrespective of the guidance law it follows. The control profile in \Cref{fig:L_latax_indp_aD_dvD,fig:L_linAcc_indp_aD_dvD} indicates that the asset ship and the defender jointly maneuver to satisfy all the objectives ( as the sliding surfaces in \Cref{fig:L_surface_indp_aD_dvD} approach zero). The corresponding speed profile is presented in \Cref{fig:L_Speed_indp_aD_dvD,fig:L_speedAgentwise_indp_aD_dvD}. 
\subsection{Effectiveness in the absence of certain control authority}
\begin{figure}[!ht]
    \centering
    \begin{subfigure}{0.8\linewidth}
    \centering
    \includegraphics[width=\linewidth]{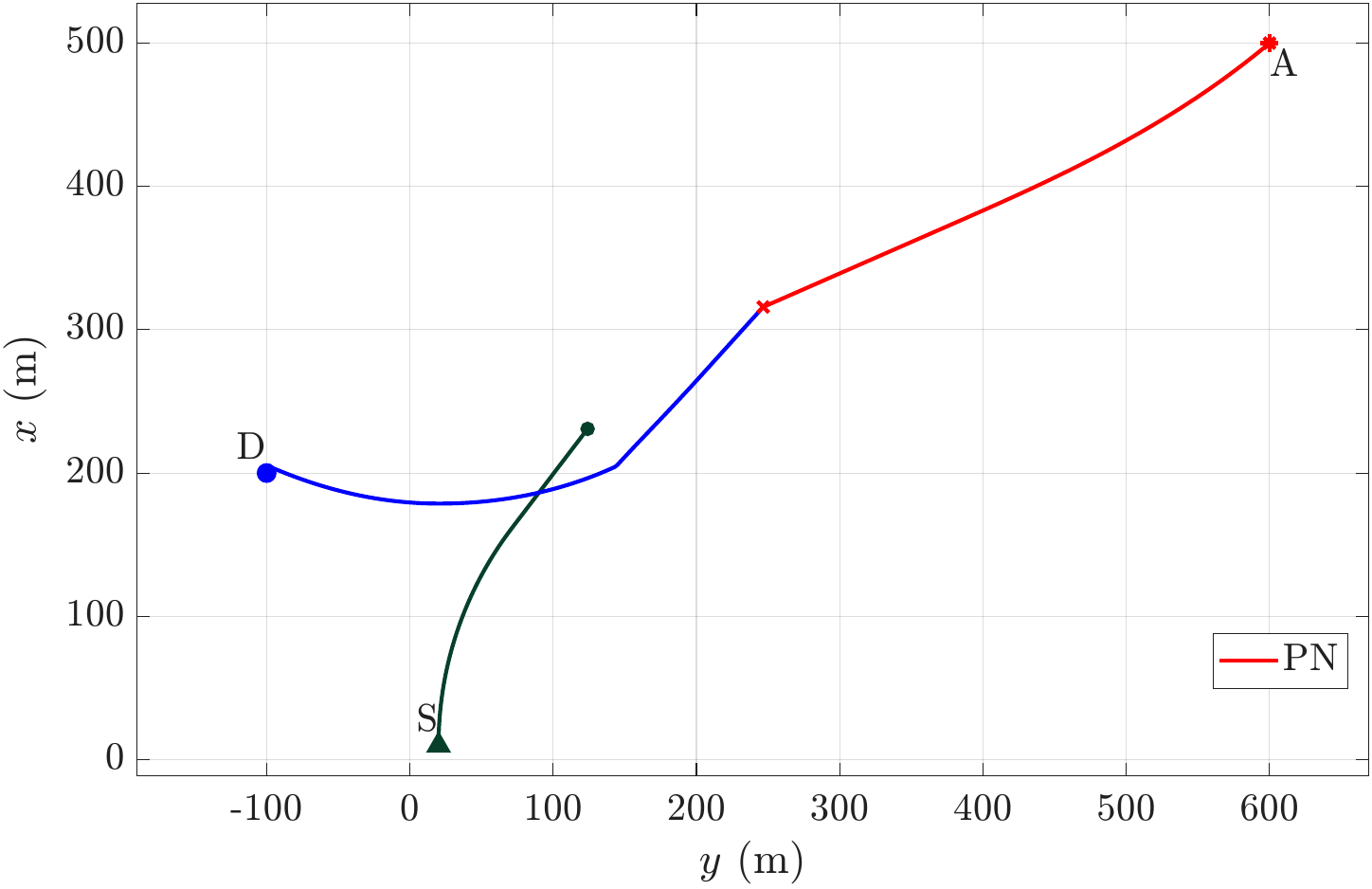}   
    \caption{Trajectories of the asset, defender, and the attacker.}
    \label{fig:PN_path_dVS_zero}
    \end{subfigure}
    \begin{subfigure}{0.8\linewidth}
    \centering
    \includegraphics[width=\linewidth]{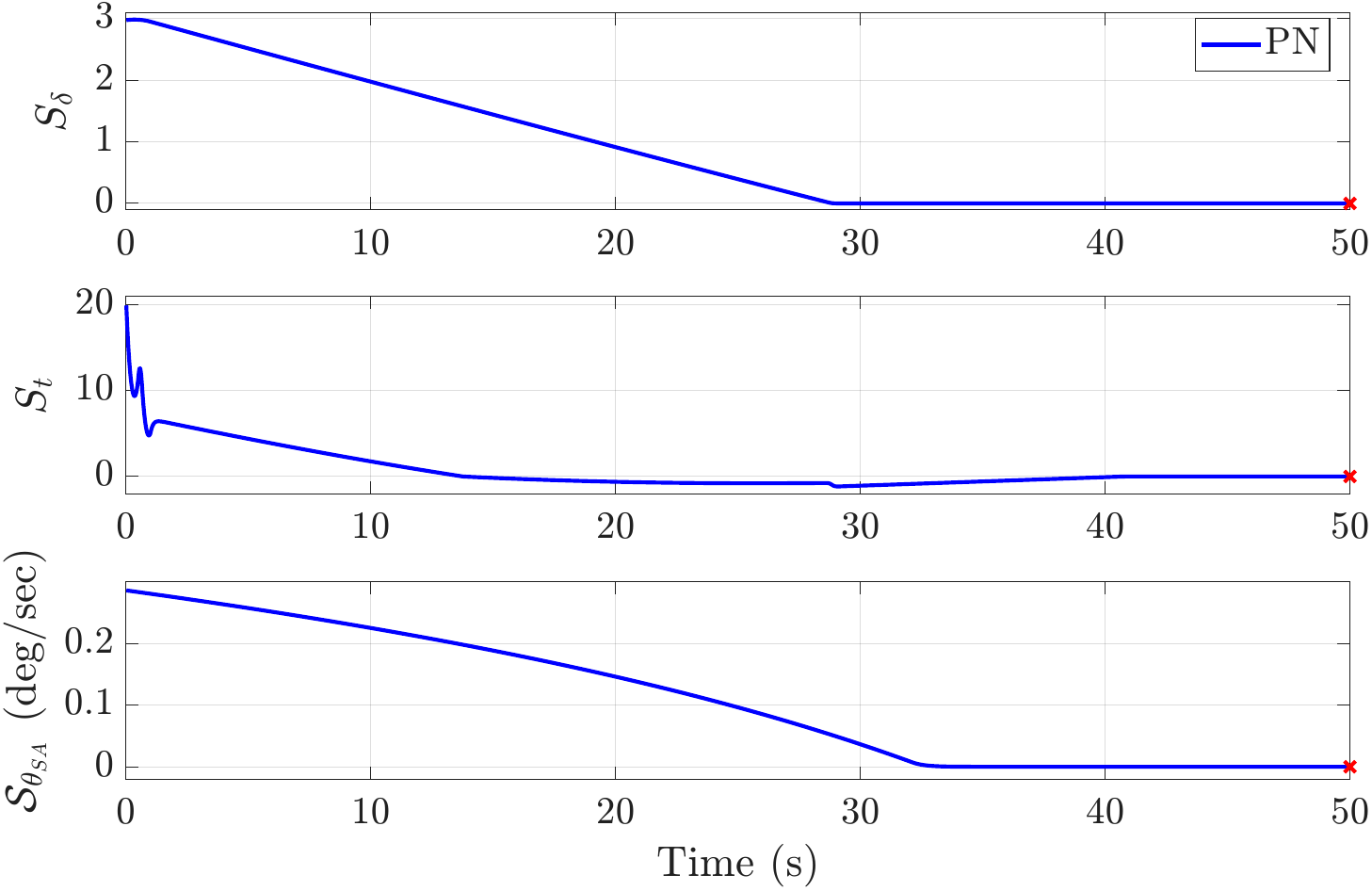}
    \caption{Sliding surfaces.}
    \label{fig:PN_surface_dVS_zero}
    \end{subfigure}
    \caption{Interception of the attacker when $\dot{V}_S=0$.}
    \label{figcase:dVS_zero}
\end{figure}

In this subsection, we further demonstrate the strategy's effectiveness using a different set of available control inputs to capture the attacker before it reaches the asset ship. For this, we again consider three cases similar to the cases as discussed in the previous subsection.
Out of the four control inputs ($\dot{V}_S,~a_S,~a_D,~\dot{V}_D$) available, we first analyze the case when $\dot{V}_S=0$, then second case is when $a_S=0$ and lastly when $\dot{V}_D=0$.
The performance corresponding to the first case, when the asset ship cannot change its speed, is depicted in \Cref{figcase:dVS_zero}. 
Even with no control over the speed of the asset, the defender and the asset ship could jointly maneuver to achieve success (refer to paths in \Cref{fig:PN_path_dVS_zero}). The sliding surface in \Cref{fig:PN_surface_dVS_zero} also converges to zero. 
\begin{figure}[!ht]
    \centering
    \begin{subfigure}{0.8\linewidth}
    \centering
    \includegraphics[width=\linewidth]{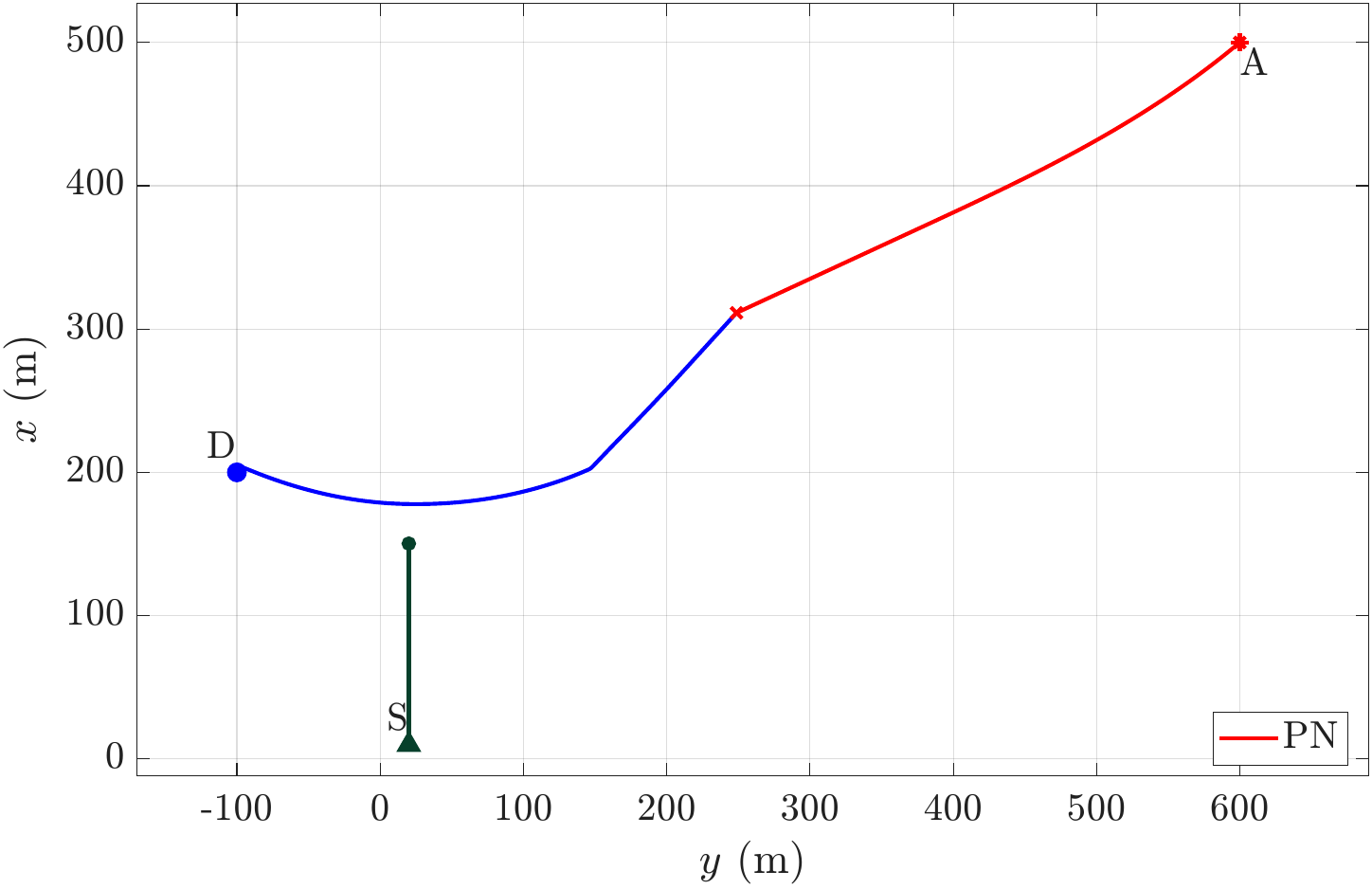}   
    \caption{Trajectories of asset, defender, and attacker.}
    \label{fig:PN_path_aS_zero}
    \end{subfigure}
    \begin{subfigure}{0.8\linewidth}
    \centering
    \includegraphics[width=\linewidth]{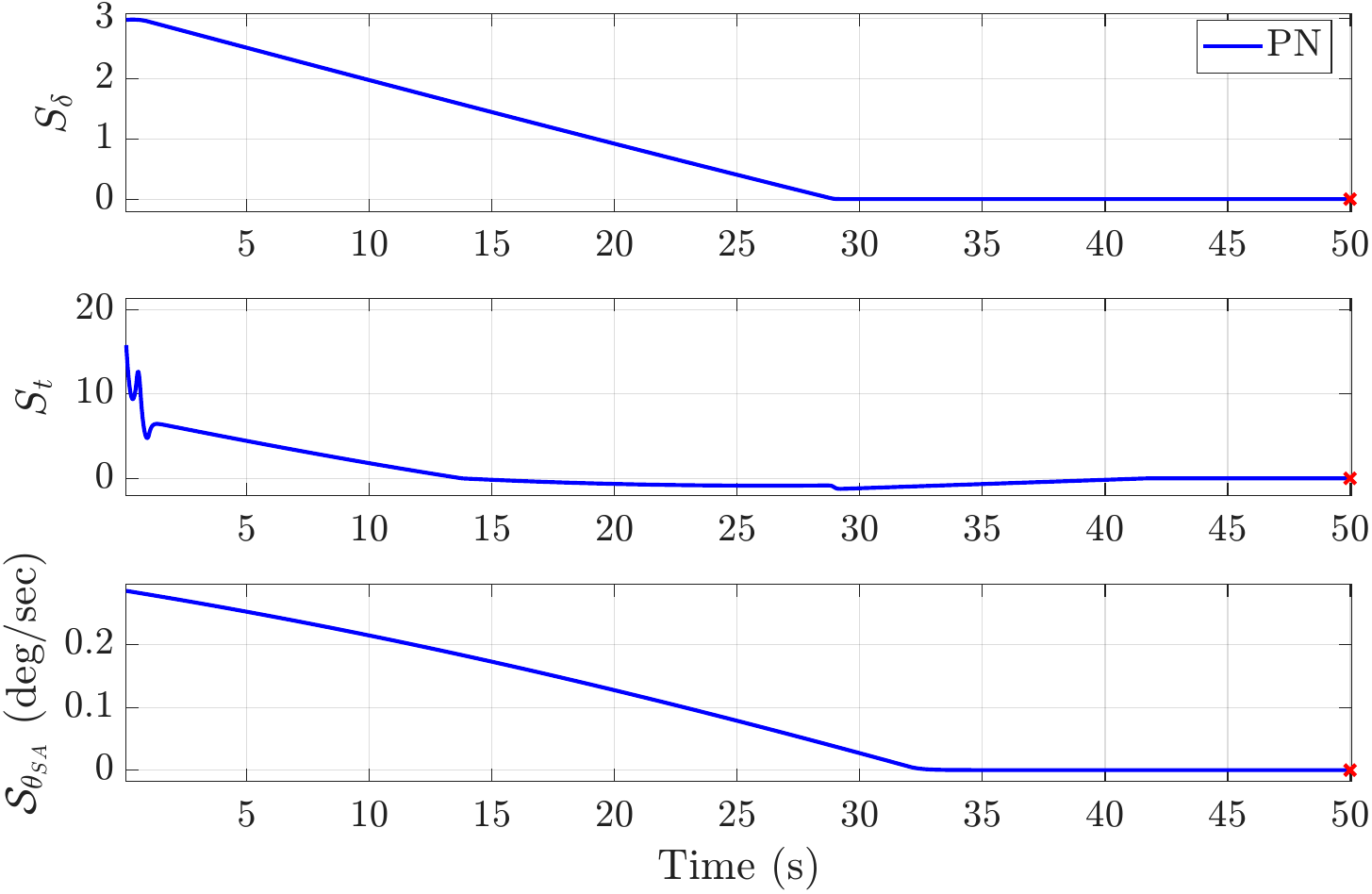}
    \caption{Sliding surfaces.}
    \label{fig:PN_surface_aS_zero}
    \end{subfigure}
    \caption{Interception of the attacker when $a_S=0$.}
    \label{figcase:aS_zero}
\end{figure}
Next, \Cref{figcase:aS_zero} demonstrates the performance of the proposed strategy for the case when the ship cannot change its direction. \Cref{fig:PN_path_aS_zero} depicts the paths of the agents when the defender successfully intercepted the attacker while the asset ship continued to move in a fixed direction. The sliding surfaces, as shown in \Cref{fig:PN_surface_aS_zero}, also converge to zero, ensuring all objectives are met. 
\begin{figure}[!ht]
    \centering
    \begin{subfigure}{0.9\linewidth}
    \centering
    \includegraphics[width=\linewidth]{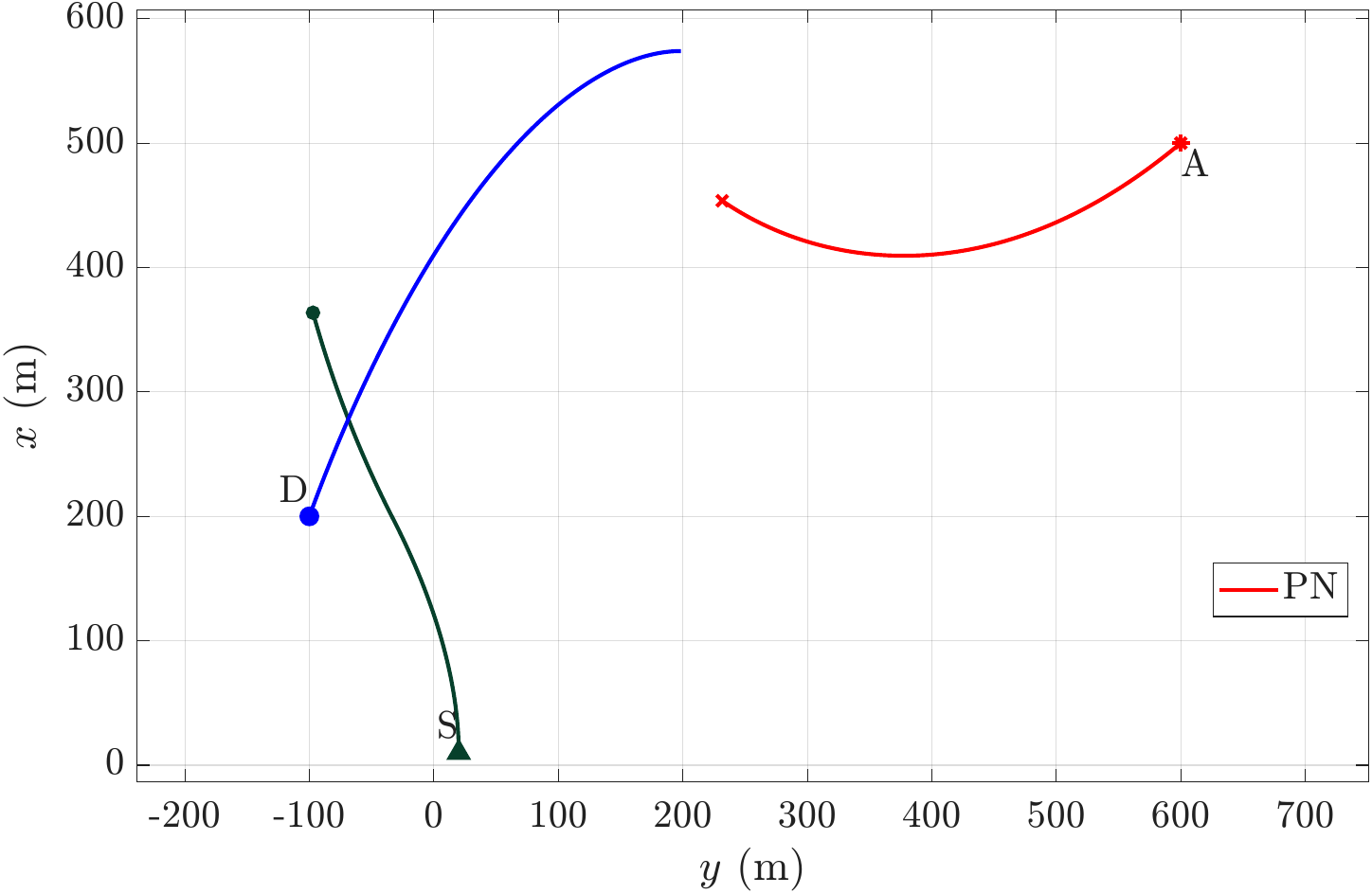}   
    \caption{Path of the asset, defender, and the attacker.}
    \label{fig:PN_path_dVD_zero}
    \end{subfigure}
    \begin{subfigure}{0.9\linewidth}
    \centering
    \includegraphics[width=\linewidth]{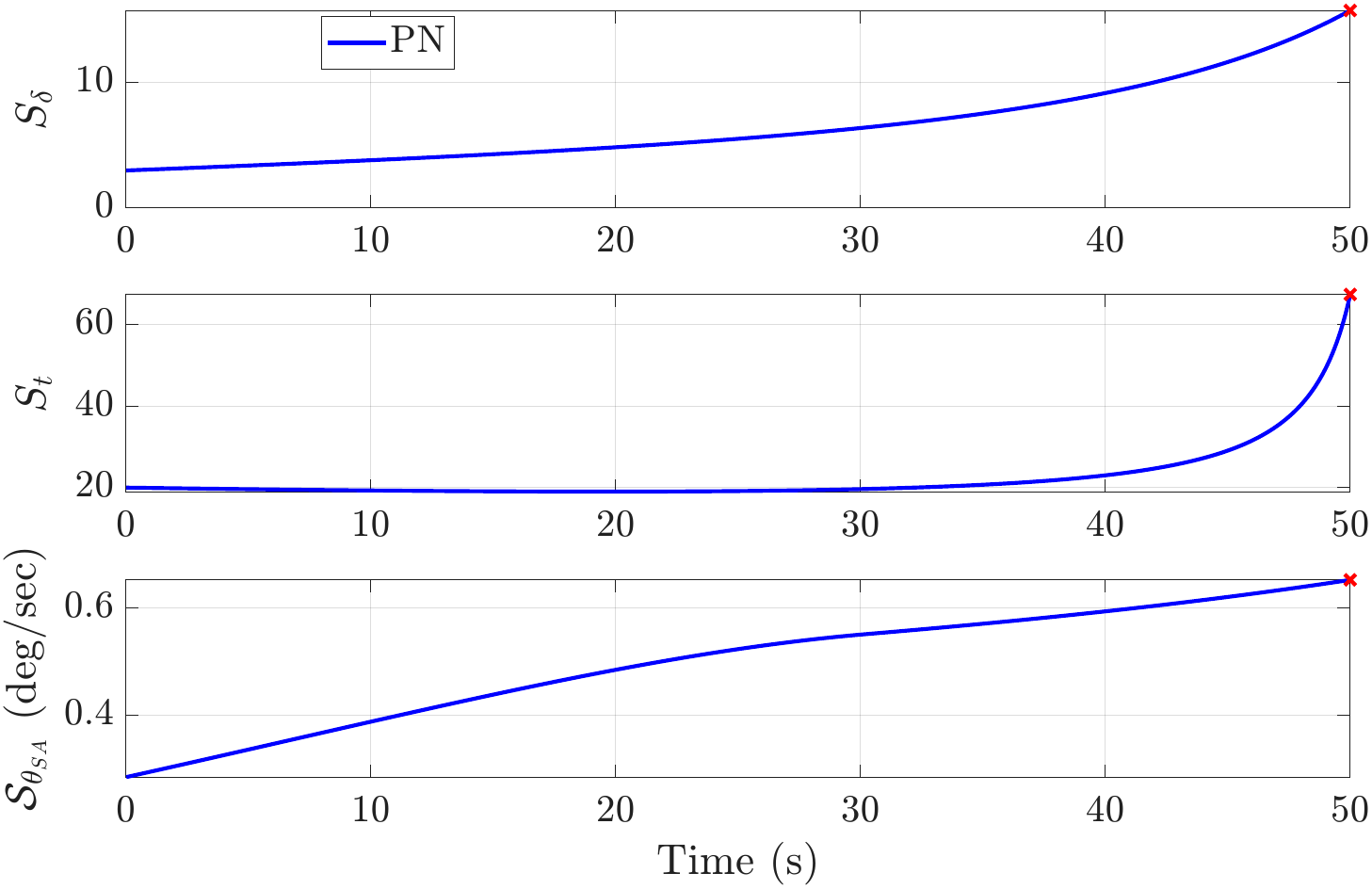}
    \caption{Sliding Surfaces.}
    \label{fig:PN_surface_dVD_zero}
    \end{subfigure}
    \caption{No interception of the attacker when $\dot{V}_D=0$.}
    \label{figcase:dVD_zero}
\end{figure}
Lastly, we consider the case in which the defender cannot control its speed (that is, $\dot{V}_D=0$). Since this case led to a loss of controllability, as indicated by the determinant of the control matrix $\mathcal{G}$.
The simulation results in \Cref{figcase:dVD_zero} are consistent. The paths of the agents shown in \Cref{fig:PN_path_dVD_zero} depict the failure in interception, and the sliding surfaces in \Cref{fig:PN_surface_dVD_zero} diverge. 

\section{Conclusion}\label{sec:conclusion}
In this paper, a cooperative asset protection strategy was proposed. The control inputs to the asset and the defender are designed simultaneously to meet three objectives: maintaining LOS-rate between the asset and the attacker, placing the defender on the LOS between the asset and the attacker, and capturing the attacker in a time-constrained manner. 
Based on numerical simulations, we concluded that the objectives were met and that the attacker was captured by the defender through a joint effort between the asset and the defender. 
In contrast to the existing methods, we accounted for the time-varying speeds of the agents involved. 
This marks a significant deviation from existing works in the interceptor guidance domain while also generalizing the constant-speed problems previously addressed. Additionally, we analyzed the proposed strategy under varying levels of control authority for the asset and the defender. The joint maneuvers of the asset ship and the defender could achieve interception even when, in one case, the speed control for the asset ship was absent, and in the second case, when the heading control of the asset ship was unavailable. Furthermore, consideration of varying agent speeds presents a more realistic scenario suitable for practical situations where a strategic collaboration can be achieved by the team of the asset and the defender to neutralize the incoming threat.

\bibliography{references}

@ARTICLE{10.1109/TAES.2020.3046328,
  author={Kumar, Shashi Ranjan and Mukherjee, Dwaipayan},
  journal={IEEE Transactions on Aerospace and Electronic Systems}, 
  title={Cooperative Active Aircraft Protection Guidance Using Line-of-Sight Approach}, 
  year={2021},
  volume={57},
  number={2},
  pages={957-967},
  doi={10.1109/TAES.2020.3046328}
}

@ARTICLE{10.1109/TAES.2025.3649483,
  author={Bajpai, Shivam and Sinha, Abhinav and Kumar, Shashi Ranjan},
  journal={IEEE Transactions on Aerospace and Electronic Systems}, 
  title={Cooperative Guidance for Aerial Defense in Multiagent Systems}, 
  year={2026},
  volume={62},
  number={},
  pages={3700-3710},
  doi={10.1109/TAES.2025.3649483}}

@inproceedings{10.2514/6.2010-7876,
  title={Triangle intercept guidance for aerial defense},
  author={Yamasaki, Takeshi and Balakrishnan, S},
  booktitle={AIAA Guidance, Navigation, and Control conference},
  pages={7876},
  year={2010},
  doi={10.2514/6.2010-7876}
}

@ARTICLE{10.1109/JOE.2025.3634663,
  author={Çelik, Ugurcan and Perrusquía, Adolfo},
  journal={IEEE Journal of Oceanic Engineering}, 
  title={USV Pursuit–Evasion Using a Complementary Scientific Machine Learning With Control Barrier Functions Approach}, 
  year={2026},
  volume={},
  number={},
  pages={1-13},
  doi={10.1109/JOE.2025.3634663}}

@ARTICLE{10.1109/LCSYS.2020.3041799,
  author={Sinha, Abhinav and Kumar, Shashi Ranjan and Mukherjee, Dwaipayan},
  journal={IEEE Control Systems Letters}, 
  title={Cooperative Salvo Based Active Aircraft Defense Using Impact Time Guidance}, 
  year={2021},
  volume={5},
  number={5},
  pages={1573-1578},
  doi={10.1109/LCSYS.2020.3041799}
  }

@inproceedings{10.2514/6.2021-1881,
  title={Optimal evasion in an active target defense scenario},
  author={Weintraub, Isaac E and Garcia, Eloy and Casbeer, David and Pachter, Meir},
  booktitle={AIAA Scitech 2021 Forum},
  pages={1881},
  year={2021},
  doi={10.2514/6.2021-1881}
}

@ARTICLE{10.1109/TAES.1976.308338,
  author={Boyell, Roger L.},
  journal={IEEE Transactions on Aerospace and Electronic Systems}, 
  title={Defending a Moving Target Against Missile or Torpedo Attack}, 
  year={1976},
  volume={AES-12},
  number={4},
  pages={522-526},
  doi={10.1109/TAES.1976.308338}}

@article{10.3182/20110828-6-IT-1002.02587,
  title={Guidance laws in target—missile—defender scenario with an aggressive defender},
  author={Rusnak, Ilan and Weiss, H and Hexner, Gyorgy},
  journal={IFAC Proceedings Volumes},
  volume={44},
  number={1},
  pages={9349--9354},
  year={2011},
  publisher={Elsevier},
  doi={10.3182/20110828-6-IT-1002.02587}
}

@article{10.2514/1.61832,
  title={Three-player pursuit and evasion conflict},
  author={Rubinsky, Sergey and Gutman, Shaul},
  journal={Journal of Guidance, Control, and Dynamics},
  volume={37},
  number={1},
  pages={98--110},
  year={2014},
  publisher={American Institute of Aeronautics and Astronautics},
  doi={10.2514/1.61832}
}

@article{10.2514/1.G001083,
  title={Cooperative strategies for optimal aircraft defense from an attacking missile},
  author={Garcia, Eloy and Casbeer, David W and Pachter, Meir},
  journal={Journal of Guidance, Control, and Dynamics},
  volume={38},
  number={8},
  pages={1510--1520},
  year={2015},
  publisher={American Institute of Aeronautics and Astronautics},
  doi={10.2514/1.G001083}
}

@article{10.2514/6.2012-4908,
  title={Linear quadratic optimal cooperative strategies for active aircraft protection},
  author={Prokopov, Oleg and Shima, Tal},
  journal={Journal of Guidance, Control, and Dynamics},
  volume={36},
  number={3},
  pages={753--764},
  year={2013},
  publisher={American Institute of Aeronautics and Astronautics},
  doi={10.2514/6.2012-4908}
}

@article{10.1016/j.oceaneng.2022.112742,
  title={Cooperative strategy for pursuit-evasion problem with collision avoidance},
  author={Sun, Zhiyuan and Sun, Hanbing and Li, Ping and Zou, Jin},
  journal={Ocean Engineering},
  volume={266},
  pages={112742},
  year={2022},
  publisher={Elsevier},
  doi={10.1016/j.oceaneng.2022.112742}
}

@ARTICLE{10.1109/TCYB.2019.2958548,
  author={Fang, Xu and Wang, Chen and Xie, Lihua and Chen, Jie},
  journal={IEEE Transactions on Cybernetics}, 
  title={Cooperative pursuit with multi-pursuer and one faster free-moving evader}, 
  year={2022},
  volume={52},
  number={3},
  pages={1405-1414},
  doi={10.1109/TCYB.2019.2958548}
}

@article{10.1007/s10846-022-01570-y,
  title={Three-agent time-constrained cooperative pursuit-evasion},
  author={Sinha, Abhinav and Kumar, Shashi Ranjan and Mukherjee, Dwaipayan},
  journal={Journal of Intelligent \& Robotic Systems},
  volume={104},
  number={2},
  pages={28},
  year={2022},
  publisher={Springer},
  doi={10.1007/s10846-022-01570-y}
}

@article{10.2514/1.G000659,
  title={Cooperative nonlinear guidance strategies for aircraft defense},
  author={Kumar, Shashi Ranjan and Shima, Tal},
  journal={Journal of Guidance, Control, and Dynamics},
  volume={40},
  number={1},
  pages={124--138},
  year={2017},
  publisher={American Institute of Aeronautics and Astronautics},
  doi={10.2514/1.G000659}
}

@ARTICLE{10.1109/TCNS.2025.3649094,
  author={Li, Ronghui and Gu, Nan and Wang, Dan and Peng, Zhouhua and Zhang, Weidong},
  journal={IEEE Transactions on Control of Network Systems}, 
  title={Safety-Critical Cooperative Pursuit Planning and Control of Multiple Autonomous Surface Vehicles Against a Partially Unknown Faster Evader}, 
  year={2025},
  volume={},
  number={},
  pages={1-10},
  doi={10.1109/TCNS.2025.3649094}
}

@article{10.1016/j.oceaneng.2026.124314,
author = {Fangyuan Xu and Zhouhua Peng and Nan Gu and Anqing Wang and Haoliang Wang},
title = {Safety-certified pursuit-evasion game of underactuated autonomous surface vehicles with experiments: A differential game approach with rate-tunable CBF optimization},
journal = {Ocean Engineering},
volume = {351},
pages = {124314},
year = {2026},
issn = {0029-8018},
doi = {10.1016/j.oceaneng.2026.124314},
}

@ARTICLE{ghose1994,
  author={Ghose, D.},
  journal={IEEE Transactions on Aerospace and Electronic Systems}, 
  title={True proportional navigation with maneuvering target}, 
  year={1994},
  volume={30},
  number={1},
  pages={229-237},
  doi={10.1109/7.250423}}

@article{kumar2022true,
  title={True-proportional-navigation inspired finite-time homing guidance for time constrained interception},
  author={Kumar, Shashi Ranjan and Mukherjee, Dwaipayan},
  journal={Aerospace Science and Technology},
  volume={123},
  pages={107499},
  year={2022},
  publisher={Elsevier}
}

@article{10.1007/s10846-016-0379-3,
  title={Pursuit-evasion games of high speed evader},
  author={Ramana, M V and Kothari, Mangal},
  journal={Journal of Intelligent \& Robotic Systems},
  volume={85},
  number={2},
  pages={293--306},
  year={2017},
  publisher={Springer},
  doi={10.1007/s10846-016-0379-3}
}

@ARTICLE{10.1109/TAES.2011.5751240,
  author={Li, Dongxu and Cruz, Jose B.},
  journal={IEEE Transactions on Aerospace and Electronic Systems}, 
  title={Defending an Asset: A Linear Quadratic Game Approach}, 
  year={2011},
  volume={47},
  number={2},
  pages={1026-1044},
  doi={10.1109/TAES.2011.5751240}
  }

@article{10.1016/j.automatica.2025.112629,
title = {Nature-inspired dynamic control for pursuit-evasion of robots},
journal = {Automatica},
volume = {183},
pages = {112629},
year = {2026},
issn = {0005-1098},
doi = {10.1016/j.automatica.2025.112629},
author = {Panpan Zhou and Sirui Li and Benyun Zhao and Bo Wahlberg and Xiaoming Hu}
}

@article{10.2514/1.G004068,
author = {Pachter, Meir and Von Moll, Alexander and Garcia, Eloy and Casbeer, David W. and Milutinovi\'{c}, Dejan},
title = {Two-on-One Pursuit},
journal = {Journal of Guidance, Control, and Dynamics},
volume = {42},
number = {7},
pages = {1638-1644},
year = {2019},
doi = {10.2514/1.G004068}
}

@ARTICLE{10.1109/TAES.2020.2998197,
  author={Gong, Haoran and Gong, Shengping and Li, Junfeng},
  journal={IEEE Transactions on Aerospace and Electronic Systems}, 
  title={Pursuit–Evasion Game for Satellites Based on Continuous Thrust Reachable Domain}, 
  year={2020},
  volume={56},
  number={6},
  pages={4626-4637},
  doi={10.1109/TAES.2020.2998197}
  }

@article{10.1007/s40295-025-00501-x,
  title={Orbital Three-Player Pursuit-Evasion Game},
  author={Sun, Sheng and Zhu, Hongxu and Wang, Wei},
  journal={The Journal of the Astronautical Sciences},
  volume={72},
  number={3},
  pages={22},
  year={2025},
  publisher={Springer},
  doi={10.1007/s40295-025-00501-x}
}

\end{document}